\newcommand{\heII}{\mbox{He~{\sc ii}}}

\newcommand{\cIV}{\mbox{C~{\sc iv}}}

\newcommand{\nV}{\mbox{N~{\sc v}}}

\newcommand{\Lstar}{\mbox{$L^\star$}}

\newcommand{\diso}{$d_{\rm iso}$}

\documentclass[a4paper,fleqn,usenatbib,twocolumn]{aastex631}
\usepackage{newtxtext,newtxmath}
\usepackage[T1]{fontenc}
\usepackage{xcolor}
\usepackage{longtable}
\usepackage{tablefootnote}
\usepackage{graphicx}	
\usepackage{amsmath}	
\usepackage{multirow}

\received{\today}
\revised{XXX}
\accepted{XXX}
\submitjournal{ApJ}

\usepackage{comment}
\usepackage{threeparttable}
\usepackage{booktabs}
\graphicspath{{plots/}}
\shorttitle{A variability census of supermassive black holes across the Universe}
\shortauthors{Cammelli et al.} 
\begin{document}

\title{Glimmers in the Cosmic Dawn. II. A variability census of supermassive black holes across the Universe\footnote{This research is based on observations made with the NASA/ESA \textit{Hubble} Space Telescope obtained from the Space Telescope Science Institute, which is operated by the Association of Universities for Research in Astronomy, Inc., under NASA contract NAS 5–26555. These observations are associated with programs 1563,12498,17073.}}

\correspondingauthor{Vieri Cammelli}
\email{vieri.cammelli@unimore.it}

\author[0000-0002-2070-9047]{Vieri Cammelli}
\affiliation{Department of Physics, Informatics \& Mathematics, University of Modena \& Reggio Emilia, via G. Campi 213/A, 41125, Modena, Italy}
\affiliation{Dipartimento di Fisica, Sezione di Astronomia, Università degli Studi di Trieste, via G.B. Tiepolo 11, I-34131, Trieste, Italy}
\affiliation{INAF - Osservatorio Astronomico di Trieste, via G.B. Tiepolo 11, I-34131, Trieste, Italy}
\affiliation{Department of Space, Earth \& Environment, Chalmers University of Technology, SE-412 96 Gothenburg, Sweden}
\affiliation{IFPU – Institute for Fundamental Physics of the Universe, Via Beirut 2, I-34151 Trieste, Italy}

\author[0000-0002-3389-9142]{Jonathan C. Tan}
\affiliation{Department of Space, Earth \& Environment, Chalmers University of Technology, SE-412 96 Gothenburg, Sweden}
\affiliation{Department of Astronomy, University of Virginia, Charlottesville, VA 22904, USA}

\author[0000-0001-9136-3701]{Alice R. Young}
\affiliation{Stockholm University, Department of Astronomy and Oskar Klein Centre for Cosmoparticle Physics, AlbaNova University Centre, SE-10691 Stockholm, Sweden}

\author[0000-0001-8587-218X]{Matthew J. Hayes}
\affiliation{Stockholm University, Department of Astronomy and Oskar Klein Centre for Cosmoparticle Physics, AlbaNova University Centre, SE-10691 Stockholm, Sweden}

\author[0000-0002-6260-1165]{Jasbir Singh}
\affiliation{INAF – Astronomical Observatory of Brera, via Brera 28, I-20121 Milan, Italy}

\author[0000-0001-7782-7071]{Richard S. Ellis}
\affiliation{Department of Physics and Astronomy, University College London, Gower Street, London WC1E 6BT, UK}

\author[0000-0001-5333-9970]{Aayush Saxena}
\affiliation{Department of Physics, University of Oxford, Denys Wilkinson Building, Keble Road, Oxford OX1 3RH, UK}

\author[0000-0001-7459-6335]{Nicolas Laporte}
\affiliation{Aix Marseille Université, CNRS, CNES, LAM (Laboratoire d’Astrophysique de Marseille), UMR 7326, 13388 Marseille, France}

\author[0000-0003-2083-7564]{Pierluigi Monaco}
\affiliation{Dipartimento di Fisica, Sezione di Astronomia, Università degli Studi di Trieste, via G.B. Tiepolo 11, I-34131, Trieste, Italy}
\affiliation{INAF - Osservatorio Astronomico di Trieste, via G.B. Tiepolo 11, I-34131, Trieste, Italy}
\affiliation{IFPU – Institute for Fundamental Physics of the Universe, Via Beirut 2, I-34151 Trieste, Italy}
\affiliation{INFN, Sezione di Trieste, Via Valerio 2, I-34127 Trieste, Italy}

\author[0000-0002-9642-7193]{Benjamin W. Keller}
\affiliation{Department of Physics and Materials Science, University of Memphis, 3720 Alumni Avenue, Memphis, TN 38152, USA}


\begin{abstract}
Understanding the origin and evolution of supermassive black holes (SMBH) stands as one of the most important challenges in astrophysics and cosmology, with little current theoretical consensus. Improved observational constraints on the cosmological evolution of SMBH demographics are needed.
Here we report results of a search via photometric variability for SMBHs appearing as active galactic nuclei (AGN) in the cosmological volume defined by the Hubble Ultra Deep Field (HUDF).
This work includes particular focus on a new observation carried out in 2023 with the \textit{Hubble Space Telescope (HST)} using the WFC3/IR/F140W, which is compared directly to equivalent data taken 11 years earlier in 2012. Two earlier pairs of observations from 2009 to 2012 with WFC3/IR/F105W and WFC3/IR/F160W are also analysed.
We identify 521, 188, and 109 AGN candidates as nuclear sources that exhibit photometric variability at a level of 2, 2.5 and 3~$\sigma$ in at least one filter. This sample includes 13, 3, and 2 AGN candidates at redshifts $z>6$, when the Universe was $\lesssim900$~Myr old. 
After variability and luminosity function (down to $M_{\rm UV}=-17\:$mag) completeness corrections, we estimate the co-moving number density of SMBHs, $n_{\rm SMBH}(z)$. At $z \gtrsim 6$, $n_{\rm SMBH}\gtrsim 6\times10^{-3}\:{\rm cMpc^{-3}}$. At low-$z$ our observations are sensitive to AGN fainter than $M_{\rm UV}=-17 \:$mag, and we estimate $n_{\rm SMBH}\gtrsim 10^{-2}\:{\rm cMpc^{-3}}$. We discuss how these results place strong constraints on a variety of SMBH seeding theories. 
\end{abstract}

\keywords{cosmology: reionization -- galaxies: evolution -- galaxies:
high-redshift -- galaxies:active galaxies}

\section{Introduction}\label{sec:intro}

Understanding the origin of supermassive black holes (SMBHs) is one of the most important unsolved problems of astrophysics and cosmology. These SMBHs reside in the nuclei of most large galaxies and are typically detected when undergoing accretion and appearing as active galactic nuclei (AGN). Feedback from SMBHs/AGN may play crucial roles in galaxy assembly \citep[e.g.,][]{Kormendy13, denBrok15, Graham2016}. 

The overall abundance, e.g., as measured by their co-moving number density $n_{\rm SMBH}$, of SMBH across cosmic time is of crucial importance for understanding their formation and evolution. An approximate estimate of SMBH abundance in the local Universe is $n_{\rm SMBH}\sim 4.6\times 10^{-3} \:{\rm cMpc}^{-3}$ \citep[][]{Banik.2019}, which assumes that every $\sim L_*/3$ galaxy hosts a SMBH, where $L_*$ is the characteristic luminosity. Integrating the SMBH mass function of \citet{Vika.2009} yields a similar estimate of $\sim 8.8\times 10^{-3} \:{\rm cMpc}^{-3}$. However, an undetected population of fainter SMBHs would increase these numbers. The lowest mass examples of known SMBHs have masses just below $\sim 10^5\:M_\odot$. For example, the study of \citet{2024Natur.631..285H} implies a black hole mass in the range $\sim (2-5)\times10^4\:M_\odot$ in $\omega$ Cen, which is the stripped nucleus of a dwarf galaxy accreted by the Milky Way. The mass function of SMBHs derived from the modelling of tidal disruption events by \citet{2024arXiv241017087M} extends down to $\sim 4\times 10^4\:M_\odot$. Furthermore, from this work there are hints of a turnover in the mass function on scales of $\sim 4\times 10^5\:M_\odot$. This, together with the general lack of good examples of intermediate-mass black holes (IMBHs) in the mass range of $\sim 10^3 - 10^4\:M_\odot$, suggests that the initial seed masses for SMBHs may be already in or near the ``supermassive'' regime of $\sim 10^5\:M_\odot$.

SMBHs with masses $\sim 10^9\: M_{\odot}$ have been found in the early universe at $z\geq7$ \citep[e.g.,][]{Mortlock13, Banados.2018,Wang21}. The \textit{James Webb Space Telescope (JWST)} has accelerated claims for high-$z$ SMBHs, including lower-mass ($\sim 10^7 - 10^8 M_{\odot}$) examples out to higher redshifts ($z\sim 10$) \citep[e.g.,][]{Harikane23, Matthee.2023, Greene23, Maiolino23b, Bogdan24, Kokorev.2024}. Assuming Eddington-limited growth, these results also imply that at least some SMBHs form early with seed masses in the supermassive regime with $\sim 10^5 M_{\odot}$ \citep[e.g.,][]{Wang21}. 

There are several proposed theories for how SMBHs form. Much attention has focused on monolithic ``direct collapse'' in metal-free, irradiated, relatively massive ($\sim 10^8\:M_\odot$), atomically-cooled halos \citep[e.g.,][]{HaehneltRees93,Begelman06,Chon16,Wise.2019}. However, the conditions required for this mechanism appear to be quite rare. For example, \citet{Chon16} found only two candidate direct collapse SMBHs in their simulation of a $\sim (30\:{\rm cMpc})^3$ volume, implying $n_{\rm SMBH}\sim 9\times10^{-5}\:{\rm cMpc}^{-3}$. In the radiation-hydro simulation of \citet{Wise.2019}, an even smaller global number density of direct collapse SMBH seeds was inferred of $n_{\rm SMBH}\sim 10^{-7}-10^{-6}\:{\rm cMpc}^{-3}$. A variant of the direct collapse scenario in which strong turbulence induced by fast converging flows supports gas in the halo against fragmentation has been proposed by \citet{2022Natur.607...48L}. However, again, the conditions for such halos appear to be very rare, yielding $n_{\rm SMBH}\lesssim 8\times 10^{-7}\:{\rm cMpc}^{-3}$.

An alternative theory of SMBH formation is based on runaway mergers of stars in dense star clusters \citep[e.g.,][]{2006MNRAS.368..141F,2023MNRAS.521.3972S}. However, the stellar densities required to produce stars that would be progenitors of even IMBHs, i.e., with $\sim 10^3\:M_\odot$, are almost never seen in local examples of young, massive star clusters \citep{2014prpl.conf..149T}. Thus, while it is difficult to predict the occurrence rate of this mechanism in cosmological volumes, it appears to require very rare, specialized conditions. Furthermore, there is very little evidence for clear examples of IMBHs with masses $\sim 10^3 - 10^4\:M_\odot$, which would be predicted to be much more common than the SMBHs that form via this mechanism.

Given these theoretical uncertainties, many implementations of SMBH seeding in numerical simulations have utilized simple threshold conditions. For example, in the Illustris project simulations \citep{Vogelsberger.2014} a Halo Mass Threshold (HMT) condition has been used \citep[based on, e.g., ][]{Sijacki07, DiMatteo08}: when a dark matter halo exceeds $7\times10^{10}\:M_\odot$, then it is seeded with a SMBH of mass $\sim 10^5-10^6\:M_\odot$. 
A natural feature of these models is that SMBH formation occurs relatively late in the universe, since it takes time for these massive halos to develop. For example, in the fiducial Illustris model, the comoving number density increases from $n_{\rm SMBH}\sim 10^{-5}\:{\rm cMpc}^{-3}$ at $z=10$ to $\sim 4\times 10^{-3}\:{\rm cMpc}^{-3}$ by $z = 5$. It continues to rise towards lower redshifts, asymptotically approaching $\sim 2\times 10^{-2}\:{\rm cMpc}^{-3}$ by $z=0$. 

In another approach, SMBH seeding recipes based on threshold conditions in the gas have been implemented. For instance, in the Romulus25 simulation suite \citep{Tremmel17} of $(25\:{\rm cMpc})^3$ volumes SMBH seeds of $10^6\:M_\odot$ are created in gas that has relatively low metallicity ($Z<3\times 10^{-4}$), a H nuclei number density $n_{\rm H}>3\:{\rm cm}^{-3}$, and a temperature between 9,500~K and 10,000~K. These conditions are designed to seed SMBHs in gas that is collapsing relatively quickly, but still having low cooling rates. However, it should be noted that these thresholds, especially of density, describe conditions that are very far from those needed to resolve the detailed processes expected to be occurring in supermassive star and/or dense star cluster formation leading to SMBHs.


An alternative theoretical model of SMBH seeding in cosmological volumes is based on the formation of supermassive Pop III.1 stars in locally isolated dark matter minihalos \citep{Banik.2019,Singh.2023,Cammelli24.MNRAS2} \citep[see][for a review]{2024arXiv241201828T}. It relies on the physical mechanism of dark matter annihilation heating to change the structure of the protostar \citep{Spolyar.2008,2008IAUS..255...24T,Natarajan.2009,2009ApJ...693.1563F,2010ApJ...716.1397F,Rindler-Daller.2015}. In particular, if the protostar can be kept in a large, swollen state relative to that of the zero age main sequence (ZAMS) structure, then it may avoid the strong ionizing feedback that acts to limit the growth of ``standard'' Pop III stars \citep{2004ApJ...603..383T,2008ApJ...681..771M,2010AIPC.1294...34T,2011Sci...334.1250H,2014ApJ...781...60H,2014ApJ...792...32S}. The Pop III.1 model for supermassive star formation requires adiabatic contraction of the dark matter density via monolithic, relatively slow contraction of the baryons in the center of a minihalo, which has been seen in many numerical simulations of the process \citep{2002ApJ...564...23B,2002Sci...295...93A,2003ApJ...592..645Y}. The baryons cool via emission of $\rm H_2$ and HD roto-vibrational transitions, with trace amounts of these species formed in the gas phase, catalysed by the presence of residual free electrons following cosmic recombination. On the other hand, UV-irradiated metal-free minihalos, i.e., Pop III.2 sources, are thought to have elevated production of $\rm H_2$ and HD leading to fragmentation to small clusters of lower-mass, $\sim 10\:M_\odot$ stars \citep{2006MNRAS.373..128G,2006MNRAS.366..247J}, which would then not lead to significant adiabatic contraction of the dark matter density and thus a much smaller, likely negligible, impact of WIMP annihilation on the protostellar evolution. 

The above distinction between ``undisturbed'' Pop III.1 and irradiated Pop III.2 minihalos motivates the main parameter of the Pop III.1 model for SMBH seeding, i.e., the isolation distance from previous generations of stars, $d_{\rm iso}$, that is needed for a minihalo to be a Pop III.1 source. \citet{Banik.2019} modeled Pop III.1 sources in a $\sim(60\:{\rm cMpc})^3$ volume and considered a range of values of $d_{\rm iso}$ from 10 to 300~kpc (proper distance). They found that a value of $d_{\rm iso}=100\:$kpc led to a co-moving number density of SMBHs of $n_{\rm SMBH}\simeq 10^{-2}\:{\rm cMpc}^{-3}$, i.e., consistent with the observational constraints discussed above. A value of $d_{\rm iso}=50\:$kpc led to an approximately ten times larger number density of $n_{\rm SMBH}\simeq 10^{-1}\:{\rm cMpc}^{-3}$. In both cases, the bulk of the SMBH population formed before $z=20$, and then maintained a near constant number density towards lower redshifts. While the \citet{Banik.2019} study only followed the evolution down to $z=10$, \citet{Singh.2023} were able to run the simulation down to $z=0$, as well as tracking mergers between SMBH-seeded halos. As a consequence of the initially widely separated SMBH locations, i.e., typical co-moving separation of 3~Mpc in the $d_{\rm iso}=100\:$kpc case, mergers only became significant at low redshift, i.e., $z\lesssim 2$. However, even in the $d_{\rm iso}=50\:$kpc case, $n_{\rm SMBH}$ only dropped by about 30\% due to mergers by $z=0$. {\it Thus, the key prediction of the Pop III.1 model of SMBH seeding is a near constant value of $n_{\rm SMBH}$ at all redshifts up to $z\sim 20$.}





This prediction for $n_{\rm SMBH}$ motivates the need for better observational constraints at high redshifts. In this series of papers we examine AGN activity in these early, high-redshift systems ($z>6$) through \textit{photometric variability}.
Specifically we address this by re-imaging historic deep \textit{Hubble Space Telescope (HST)} field with {\it HST}.

The {\it Hubble Ultra Deep Field} (HUDF) is the deepest field for which there is a long history of exquisite {\it HST} observations. In order to probe variability to the highest redshifts, near-infrared imaging with the Wide Field Camera 3 (WFC-3) is essential and the two most relevant historic datasets are the HUDF09 (GO 11563, PI: Illingworth, 192 orbits) and HUDF12 (GO 12498, PI: Ellis, 30 orbits) imaging campaigns. 

In Paper I of this series \citep{Hayes24} we reported first results from a new observation of the HUDF that was made in 2023 in the F140W filter (GO 17073, PI: Hayes, hereafter HUDF23). The images were taken to an equivalent depth as the HUDF12 observation, and thus permit a detailed study of variability between the two epochs. An analysis of variability between the HUDF09 and HUDF12 epochs was also carried out. The main result of Paper I was the report of three high-$z$, i.e., between $z=6-7$, AGN candidates identified via variability and implication of these results for $n_{\rm SMBH}$ and thus SMBH seeding models.

Here in Paper II we present the full variability analysis of the HUDF09, HUDF12 and HUDF23 data sets, including detection of AGN candidates at various thresholds of significance and over the full redshift range of the source population. 
The plan of the paper is as follows. In \S\ref{sec:methods} we introduce the three key datasets used in our study.
For each of these we describe the processing pipeline and photometric catalogs. 
We also discuss our techniques for identifying variable candidates. \S\ref{sec:results} presents our results, including the implications for SMBH seeding models.
We present our conclusions in \S\ref{sec:conclusions}.



\section{Datasets and methods}\label{sec:methods}

Here we review the observation datasets and the methodology used in our study (see also Paper I).

\subsection{Observations and Data Reduction}\label{sec:obs}

\begin{table}
    \centering
    \begin{threeparttable}
        \caption{Observing epochs and times.\label{tab:epochs}}
        \begin{tabular}{cccc}
            \tableline
            Year & Filters & Orbits & GO\# / PI \\
            \tableline
            2009-2010  & F105W & 24 & 11563 / Illingworth \\
                    & F160W & 53 &  \\
            \tableline
            2012    & F105W & 72 & 12498 / Ellis \\
                    & F140W & 30 &  \\
                    & F160W & 26 &  \\
            \tableline
            2023    & F140W & 30 & 17073 / Hayes \\
            \tableline
        \end{tabular}
        \begin{tablenotes}
            \item This table lists the observing epochs, filters used, orbits, and the corresponding GO numbers and PIs.
        \end{tablenotes}
    \end{threeparttable}
\end{table}


The HUDF field was initially observed in the optical with ACS \citep{Beckwith.2006}, followed by key NIR mode observations with WFC3/IR during 2009-2010 under HUDF09 \citep{Bouwens.2009z10}, capturing images in three filters (F105W, F125W, and F160W) over 192 orbits. In 2012, the field was re-imaged under HUDF12 \citep{Ellis.2013}, significantly deepening the F105W and F160W exposures and adding a fourth filter, F140W, to search for Lyman break galaxies at $z\sim8$.

To search for photometric variability across all sources within the HUDF IR footprint, we re-imaged the field in September 2023 using the F140W filter, replicating the center, field orientation, and depth (30 orbits) of the HUDF12 observation. We processed the F140W image with the \texttt{calwfc3} pipeline and \texttt{astrodrizzle} \citep{DrizzlePac} software, using High-Level Science Products (HLSP) from the Mikulski Archive for Space Telescopes (MAST) as reference images. Additionally, we re-processed the F140W image from the UDF12 campaign to confirm our methods align with the depth of the HLSP image. Concurrently, we independently re-processed the F105W and F160W images from the 2009 and 2012 epochs to search for variable sources over the shorter, earlier time baseline. This setup enables variability searches across three epochs: the period from 2009 to 2012 is covered by the F105W and F160W filters, while the span from 2012 to 2023 is sampled by the F140W filter alone (see Table~\ref{tab:epochs} for details).

\subsection{Photometry}\label{sec:photometry}

For a given filter and time baseline, we run \texttt{Source Extractor} \citep{Sextractor} on both epochs. We use an rms map based on the weight map produced by \texttt{astrodrizzle} for each filter, which is itself an inverse variance image based on the input exposures that contributed to each pixel. For each detection, \texttt{Source Extractor} outputs the geometric properties of the detection ellipse, specifically WCS and pixel coordinates, and photometric properties of detected object, namely the flux, the magnitude and their respective errors estimated accordingly. 

Rather than using \texttt{Source Extractor} in the standard \textit{single mode}, we opt for the \textit{dual mode} approach. 
In contrast to Paper I, where we implemented the dual mode approach using the single epochs themselves as detection images in both directions, here we employ a common detection image for any given filter and epoch. Specifically, we use a combined stacked frame from multiple epochs for the unconvolved F105W, F140W, and F160W images, along with the original HLSP in F125W, all matched to the same WCS. This extremely deep detection image, reaching nearly 300 orbits, provides a robust estimate of galaxy morphologies and the highest precision centroiding on galaxy nuclei where variability is expected. In addition, this method provides a ``global'' list of source coordinates and apertures that remains consistent across different runs of \texttt{Source Extractor}, ensuring one-to-one correspondence for all extracted photometric catalogs. 

Instead of conducting ``extended'' galaxy photometry, such as using Kron-like or moment-centered apertures as done by \texttt{Source Extractor}, we aim to obtain photometry centered on the brightest, unresolved sources within each galaxy determined from the ultra-deep stack. To achieve this, we rely on the coordinates of the barycenter (centroid) pixels reported by \texttt{Source Extractor} and the background images it generates during its background subtraction process. At these coordinates, we perform aperture photometry within 4-pixel diameter ($0\farcs 26$) apertures, applying local background subtraction to exclude local (non-compact) galaxy light. 


We then assemble a photometric catalog for each individual filter at each epoch by running \texttt{Source Extractor} in the dual mode. As stated above, the dual mode assures us that a possible photometric variability, if any, comes from the same region of the sky, once every frame has been co-aligned with respect to the HLSP. Hence, for each image, we obtain photometry of each galaxy nucleus, where AGN variability is expected. We consider that any off-centre variability would likely be due to supernovae (SNe). We defer analysis of such variability to a future paper in this series.


We then compare these local aperture magnitudes in each image relative to the photometric uncertainty of each source. Additionally, the drizzling process causes artificially underestimated uncertainties due to the correlation of signal between adjacent pixels. We account for this effect by multiplying the uncertainties by a correction factor, described below, which is assumed to be constant over the processed area \citep[e.g.,][]{Casertano2000, Fruchter2002}. It is worth emphasising that our photometric measurements do not aim to an accurate estimation in absolute terms, rather to a comparison between two different epochs in terms of relative variation. 


\subsection{Identifying Variable Sources}\label{sec:calib_err}

Following Paper I, we identify variables in each matched pair of filters using two techniques: comparing the nuclear/central photometry of galaxies at different epochs and detecting residual sources in pair-subtracted images in any given filter. Regions near the image edges are excluded due to the dithering pattern causing excess noise and we focus solely on a central region of the HUDF covering 123\arcsec$\times$139\arcsec. 

\subsubsection{Photometric Variables}

The photometric comparison method follows a similar approach to that of \citet{OBrien.2024}. We first correct for a systematic offset of approximately 0.01 magnitudes that is observed at all magnitudes, likely due to slightly imperfect zero points in images taken many years apart as well as telescope expansion and/or breathing. Under the assumption that the majority of sources in the field will not vary, we calculate the standard deviation of the $\Delta m$ distribution in 0.5~mag bins, where $\Delta m=m_1-m_2$ is the magnitude difference. The magnitude labels always refer to the time ordering of the different epochs, with ``1'' being the first visit and ``2'' the second. We then propagate the uncertainty on the magnitude difference derived from our photometric catalogs in the following way: 
\begin{equation}\label{eq:deltamag_err}
    \delta_{\Delta m} = \Big(\frac{2.5}{{\rm ln}\:10}\Big) \sqrt{\Big(\frac{\delta_{F_{\mathrm 1}}}{F_1}\Big)^2 + \Big(\frac{\delta_{F_{\mathrm 2}}}{F_2}\Big)^2}, 
\end{equation}
where $\delta_{F}$ and $F$ are the flux error and the flux as estimated by \texttt{Source Extractor}, respectively.

To calibrate our estimates of the uncertainties, we compare the observed standard deviations of the $\Delta m$ distributions in magnitude bins of $0.5\:$mag width over the range from 24 to 31 mag with the estimated mean photometric uncertainties in each bin from Eq.~\eqref{eq:deltamag_err}. 
 
We then compute the total average scale factor that globally adjusts the uncertainty reported in Eq.~\eqref{eq:deltamag_err} to match the observed $\Delta m$ standard deviation. For the F105W, F140W, and F160W filters the resulting calibration scale factors are 0.80, 0.84, and 0.79, respectively. We refer to the calibrated uncertainties as $\delta_{\Delta m,{\rm cal}}$.

It is by taking the ratio of the $\Delta m$ with the uncertainty calculated in Eq.~\ref{eq:deltamag_err} that a given source's significance of variability is ultimately assessed in a two-epoch observation in a given filter. 
We verify that this empirical approach follows the trend of the calibrated uncertainties by carrying out a linear fit to the logarithm of the calibrated $1\sigma$ uncertainty estimates in each 0.5 mag bin (see Fig.~\ref{fig:delta_mag_calib}). We consider three thresholds of significance for variability: $2\sigma$; $2.5\sigma$; and $3\sigma$. For a source to be classified as a candidate variable, it must have a $\sigma_{\Delta m}\equiv \Delta m/\delta_{\Delta m,{\rm cal}} \ge 2$ in at least one filter.  

\subsubsection{Difference Imaging Variables}

As a complement to variability identified via aperture photometry, we also generate pair-subtracted images for each filter. From these images we may identify additional variable sources, including sources offset from galactic nuclei, that were not detected as variables via aperture photometry.
Also, for extended sources we are able to assess the morphology of photometric variables. In particular, single AGN are expected to appear as compact, unresolved variable sources. 
Then images taken in the same filters were directly subtracted, resulting in six ``difference images''. We then applied \texttt{Source Extractor} to these difference images and effective gains were re-calculated using the difference of the squared exposure times multiplied by the CCD gain to account for the increased sky noise in the subtracted images, which mimics a shallower integration time. Difference image selected sources, as we describe below, are relatively few in number, but are considered to have high significance and for the purposes of counting statistics are included in the $3\sigma$ significance photometric sample.

Finally, to prevent possible image artefacts and other forms of interlopers contaminating our sample, we visually inspect every source identified as variable in both methods. Possible interlopers include moving objects, noisy pixels, saturated sources, and telescope diffraction spikes. SNe which occur between the two epochs (either between 2009 and 2012 or between 2012 and 2023) are removed from the sample since there will be no detection in the first epoch and a bright, obvious detection in the second. For this visual inspection, along with pair-subtracted images, we check the single epoch frames in all the three filters. Furthermore, we rule out the possibility of low redshift interlopers or artefacts by examining cutouts in the optical and in the infrared by leveraging available observations with HST and JWST, respectively (see Fig.~\ref{fig:cutouts} for the used criteria). Sources identified as image artefacts or interlopers are excluded from the variable source sample.

\subsubsection{Redshift determination}\label{subsec:z_sources}

We cross-correlate each analyzed source with positions from known redshift catalogs, including spectroscopic data from VLT/MUSE \citep{Bacon.2023}, the JADES GTO program for spectroscopic redshifts using NIRSpec \citep{Bunker.2023}, photometric redshifts from NIRCam + {\it HST} \citep{Rieke.2023}, and the Ultraviolet UDF photometric catalog \citep{Rafelski.2015}. Spectroscopic estimates, when available, are given preference over photometric ones. We note that the average photometric redshift uncertainty is of the order of $\lesssim0.2$, which justifies our choice of unit width bins when considering the overall redshift distribution (see nest Section).

\section{Results}\label{sec:results}

\begin{figure*}
    \centering
        \hspace{-0.5cm}
        \includegraphics[width=0.9\textwidth]{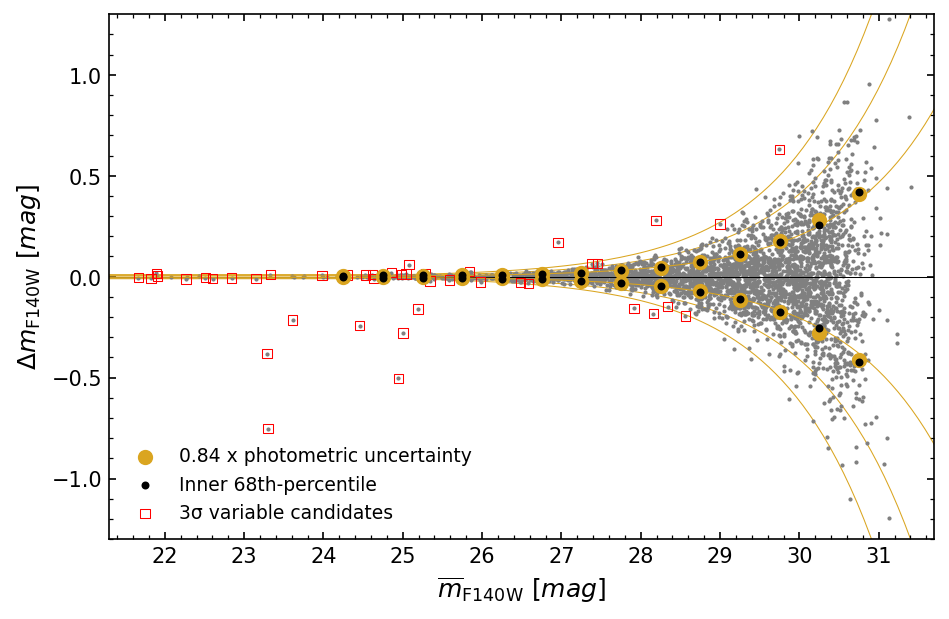}
        \includegraphics[width=0.88\textwidth]{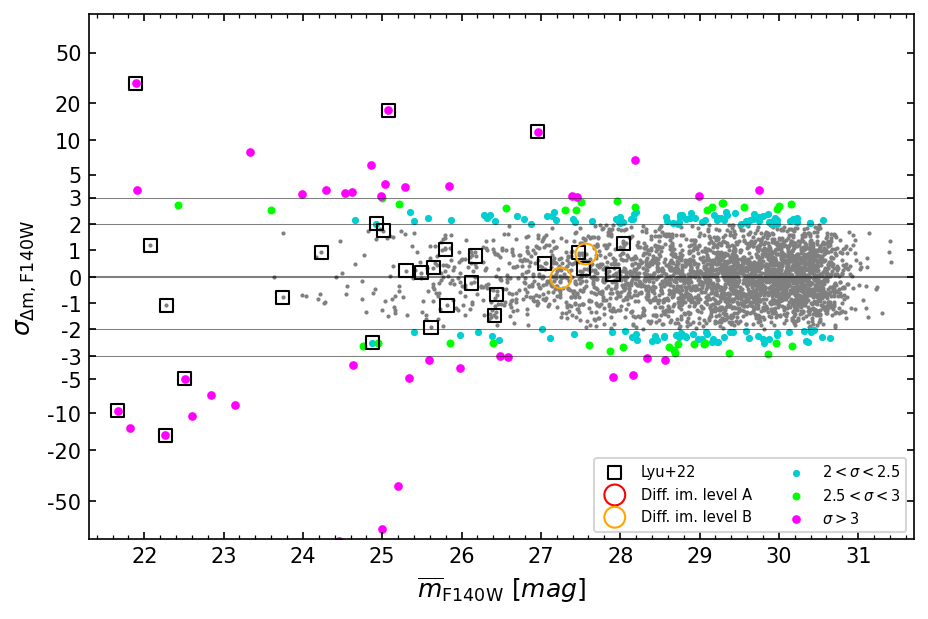}
        \caption{{\it (a) Top panel:}
        Photometric variability of HUDF sources in the F140W filter from 2012 to 2023, $\Delta m_{\rm F140W}$, versus mean magnitude, $\bar{m}_{\rm F140W}$ (gray points). 
        Black circles show the intrinsic $1\sigma$ scatter of the data measured in bins of 0.5 mag width. Yellow circles depict the calibrated (by a factor of 0.84) photometric $1\sigma$ uncertainties in each bin (see text).
        The inner pair of yellow lines are fits to these calibrated uncertainties. The next sets of yellow lines show $2\sigma$ and $3\sigma$ uncertainties, scaled from the $1\sigma$ fit.
        Red squares highlight sources that are estimated to be $\geq3\sigma$ variables.
        {\it (b): Bottom panel:} As (a), but now showing the variability significance metric, $\sigma_{\Delta m, {\rm F140W}}$, versus $\bar{m}_{\rm F140W}$. A linear scale is used for $|\sigma_{\Delta m, {\rm F140W}}|\leq3$ and a logarithmic scale elsewhere. The cross-matched known AGNs in the HUDF \citep{Lyu.2022} are marked by open squares. Variable sources identified only by difference imaging are shown in open circles. 
        }
        \label{fig:delta_mag_calib}
\end{figure*}

\subsection{Detected Variable Sources}

Figure~\ref{fig:delta_mag_calib}a shows the photometric variability ($\Delta m = m_{\rm 1}-m_{\rm 2}$) of galactic nuclei sources in the HUDF in the F140W filter from 2012 (epoch 1) to 2023 (epoch 2), i.e., an 11-year observer frame time baseline, as a function of mean magnitude, $\bar{m}$, over the range from about 21 to 32 mag. A linear fit to the logarithm of the calibrated $1\sigma$ uncertainty estimates in 0.5 mag bins (yellow circles) is shown by the inner pair of yellow lines. Note that the yellow circles represent the mean photometric uncertainties as provided by \texttt{Source Extractor} multiplied by an overall scale factor (0.84 for F140W) in order to match the inner 68th percentile of the photometric variation (black dots). This factor accounts for systematic uncertainties in the estimate of the photometric uncertainties, 
e.g., due to non-constant PSF and correlated pixel noise.  The fitted lines are then scaled by factors of two and three to show estimates of the $2\sigma$ and $3\sigma$ uncertainties. Sources that have values of $|\Delta m_{\rm F140W}|$ that are greater than $3\sigma$ have been highlighted. Note that for the faintest sources with $m_{\rm F140W}\gtrsim 30\:$mag this corresponds to a variation of $\gtrsim0.6$~mag.

Figure~\ref{fig:delta_mag_calib}b shows the dimensionless variability metric $\sigma_{\Delta m,{\rm F140W}}$ of each source versus mean magnitude. A linear scale is used for the inner range from $-3$ to $+3$ and then a logarithmic scaling at higher absolute values. The limits of $\sigma_{\Delta m,{\rm F140W}}=2, 2.5, 3$ are highlighted, showing the different levels of significance that we consider. From these data we identify 191, 67 and 39 sources that are $>2\sigma$, $>2.5\sigma$, and $>3\sigma$ variables in the F140W observations (see Table~\ref{tab:lyu}). 

In Figure~\ref{fig:delta_mag_calib}b we also highlight the cross-match with the 31 previously known AGN from the sample of \citet{Lyu.2022}, with these sources having $\bar{m}_{\rm F140W}\gtrsim 28\:$mag. We see that four of these AGN have been detected as variable sources with $\sigma_{\Delta m}>3$, still four with $\sigma_{\Delta m}>2.5$, and five with $\sigma_{\Delta m}>2$.
In \S\ref{sec:n_smbh} we discuss our efficiency of recovering AGN via our global analysis of three measurements of variability via the F105W, F140W and F160W observations.

The equivalent results of Figure~\ref{fig:delta_mag_calib} for the F105W and F160W observations are shown in Appendix~\ref{app:phot_var}. From these data, the equivalent numbers of variable sources from the F105W (2009-2012, i.e., 3-year time baseline) observation are 222, 92 and 43 sources at $2\sigma$, $2.5\sigma$, and $3\sigma$ variables, respectively. The F160W (2009-2012, i.e., 3-year time baseline) observation yields 185, 68 and 39 sources at $2\sigma$, $2.5\sigma$, and $3\sigma$ variables, respectively. We see that the F105W, F140W and F160W observations have similar efficiencies at detecting variable sources.

The difference imaging method identifies a further 6 variable sources, which were not detected as variable via the photometric method. In the following we consider these sources as part of the $\geq3\sigma$ sub-sample. Additionally, 3 more sources previously identified at $2\sigma$ and 1 source identified at $2.5\sigma$ are promoted to $3\sigma$ for a total of 10 sources added at this significance level.

Combining the above results for all the filters, in the end we obtain a total sample of 521, 188, $99+10=109$ sources that show variability in at least one filter at the $2\sigma$, $2.5\sigma$, and $3\sigma$ levels, respectively. Of these, 482, 173, and 95 sources have been cross-matched with measured redshift catalogues, respectively, with the redshift distribution discussed in the next sub-section. The photometric variability data for each source with a known redshift is presented in the electronic version of Table~\ref{tab:high_z}, with the displayed version here listing only the highest redshift ($z>6$) sources.

Finally, for completeness, we define the total variability significance of a source from the three observations using F105W, F140W and F160W via:
\begin{equation}\label{eq:sigma_tot}
    \sigma_{\rm TOT} = \sqrt{\sigma_{F105W}^2 + \sigma_{F140W}^2 + \sigma_{F160W}^2}.
\end{equation}
The values of $\sigma_{\rm TOT}$ are also listed in Table~\ref{tab:high_z}.
Note that, assuming independent Gaussian distributions for each filter measurement, the final $\sigma_{\rm TOT}$ follows the statistics of a $\chi$ distribution. We checked for correlation between F105W and F160W variability finding they are largely uncorrelated. This may be caused by the visits in the 2009-2010 epochs being taken up to 1 year apart (see \S\ref{tab:epochs}). The values of $\sigma_{\rm TOT}$ corresponding to significance levels of 2, 2.5 and 3$\sigma$ are 2.83, 3.30 and 3.76, respectively.
Nevertheless, for definition of our samples of AGN candidates, we will use the metrics based on single filter variability significance, as described above.

\subsection{Redshift Distribution}

\begin{figure*}
    \centering
	\includegraphics[width=.6\textwidth]{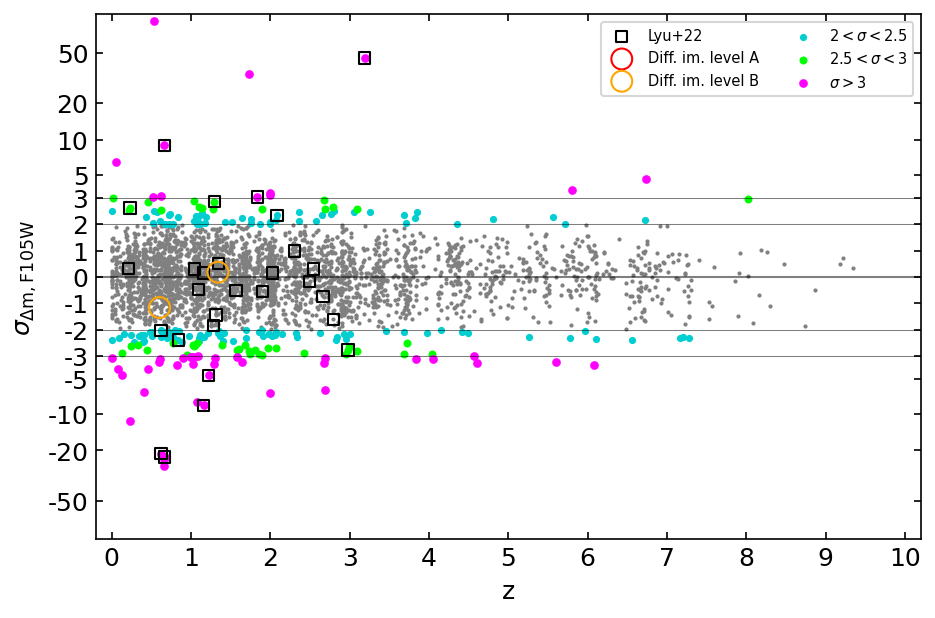}
	\includegraphics[width=.6\textwidth]{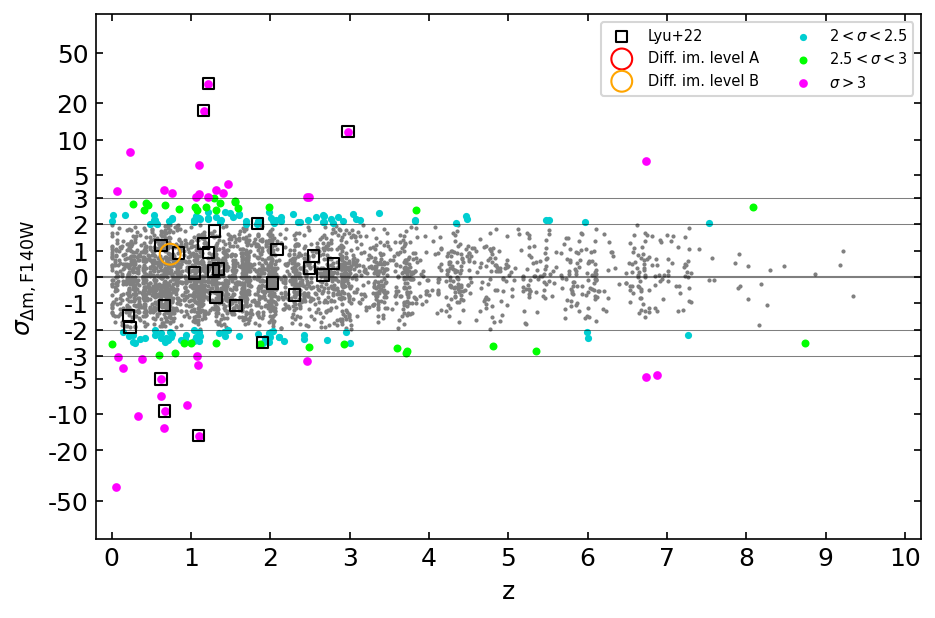}
	\includegraphics[width=.6\textwidth]{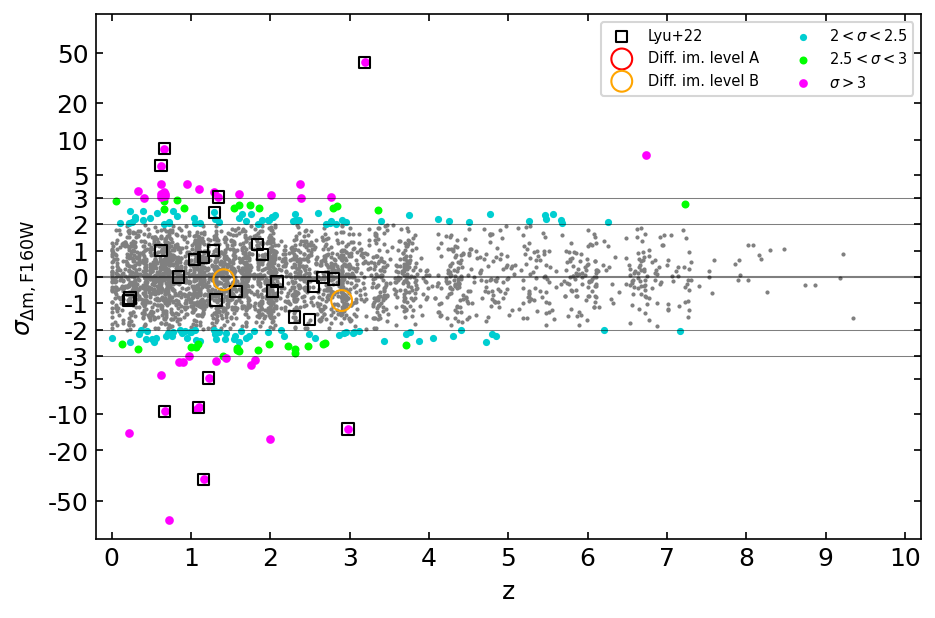}
    \caption{Variability significance metric, $\sigma_{\Delta m}$, as a function of redshift for the three filters F105W, F140W and F160W. The cross-matched known AGNs in the HUDF \citep{Lyu.2022} are marked by open squares. Sources identified by difference imaging are shown in open circles.
    }
    \label{fig:sigma_z}
\end{figure*}

\begin{figure}
	\includegraphics[width=\columnwidth]{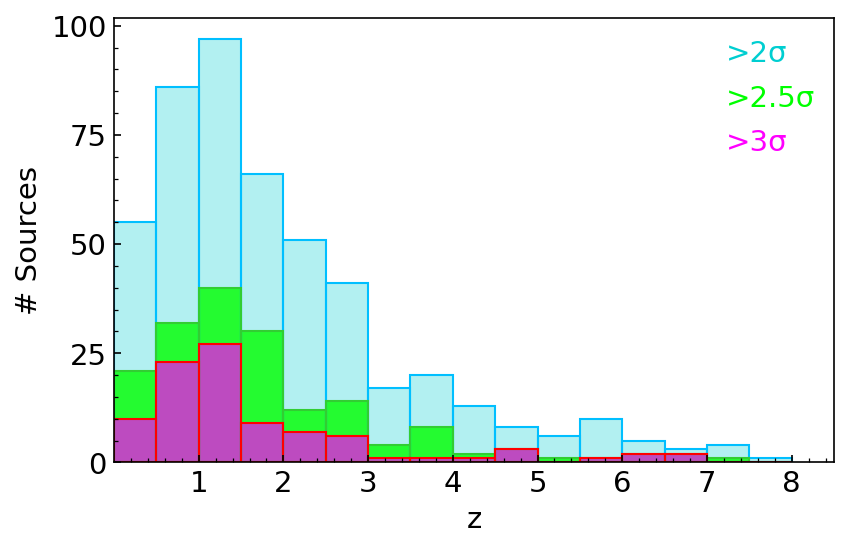}
	\includegraphics[width=\columnwidth]{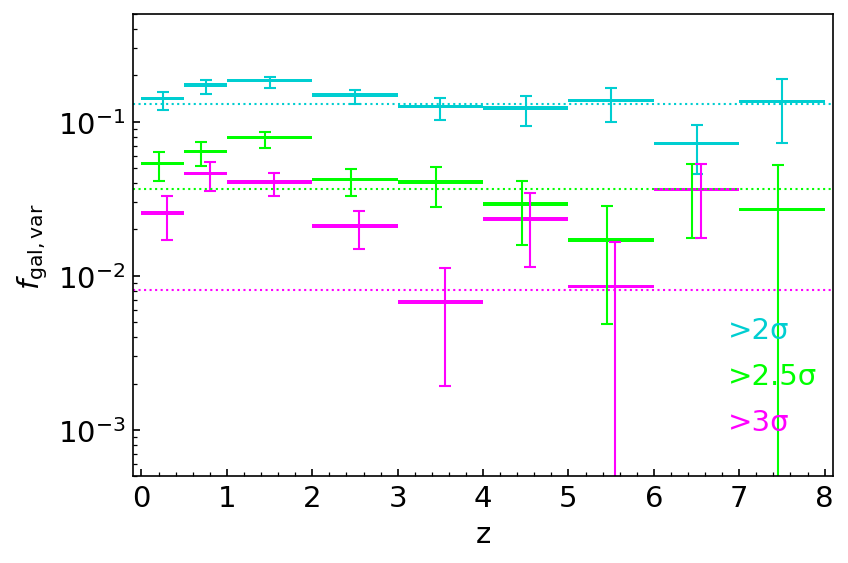}
    \caption{\textit{Top panel:} Redshift distribution of the variable candidate sources. Different colours refer to the specific significance thresholds for variability detection in units of $\sigma$. The histogram includes sources selected both via variability and via difference imaging in the 3 filters. 
    \textit{Bottom panel:} Fraction of galaxies detected as variable sources as a function of the redshift at different $\sigma$ levels (horizontal bars). The width of the bars indicates the adopted binning in redshift. Vertical error bars depict Poisson uncertainties, while dotted lines show the expected number of false positives given assuming a combined Gaussian statistic in the three filters (see text).
    }
    \label{fig:hist_z}
\end{figure}

Figure~\ref{fig:sigma_z} presents a scatter plot of $\sigma_{\Delta m}$ versus $z$ for the F105W, F140W, and F160W datasets. We see that the redshift distribution of the sources in the sample extends out to $z\sim 9.5$, with the highest redshift (at $2.5\sigma$) variability detections being at $z\sim 8$ (three sources). Among them, the highest redshift variable in our sample results at $z=7.53^{+0.46}_{-0.29}$. Table~\ref{tab:high_z} lists all the 13 variable sources that we identify with $z>6$ (note the online electronic table contains the full sample at all redshifts).

Figure~\ref{fig:hist_z}a shows the redshift histogram of our sample of detected variable sources. Different colors depict the different samples according to the significance level ($>2$, 2.5 and $3\sigma_m$) as detected in at least one filter. We see that most candidates reside at $z\lesssim4$ for all cases of significance threshold.
Figure~\ref{fig:hist_z}b shows the fraction of galaxies that are detected to have variable nuclei. This fraction appears relatively constant with redshift and is seen to be elevated compared to the false positive levels, with this enhancement relatively greater at higher values of $\sigma$. However, the absolute number of variable sources are often dominated by those found at $2\sigma$, even after subtracting off the number of expected false positives (see next sub-section).

\begin{figure*}
    \hspace*{1.1cm}
	\includegraphics[width=0.9\textwidth]{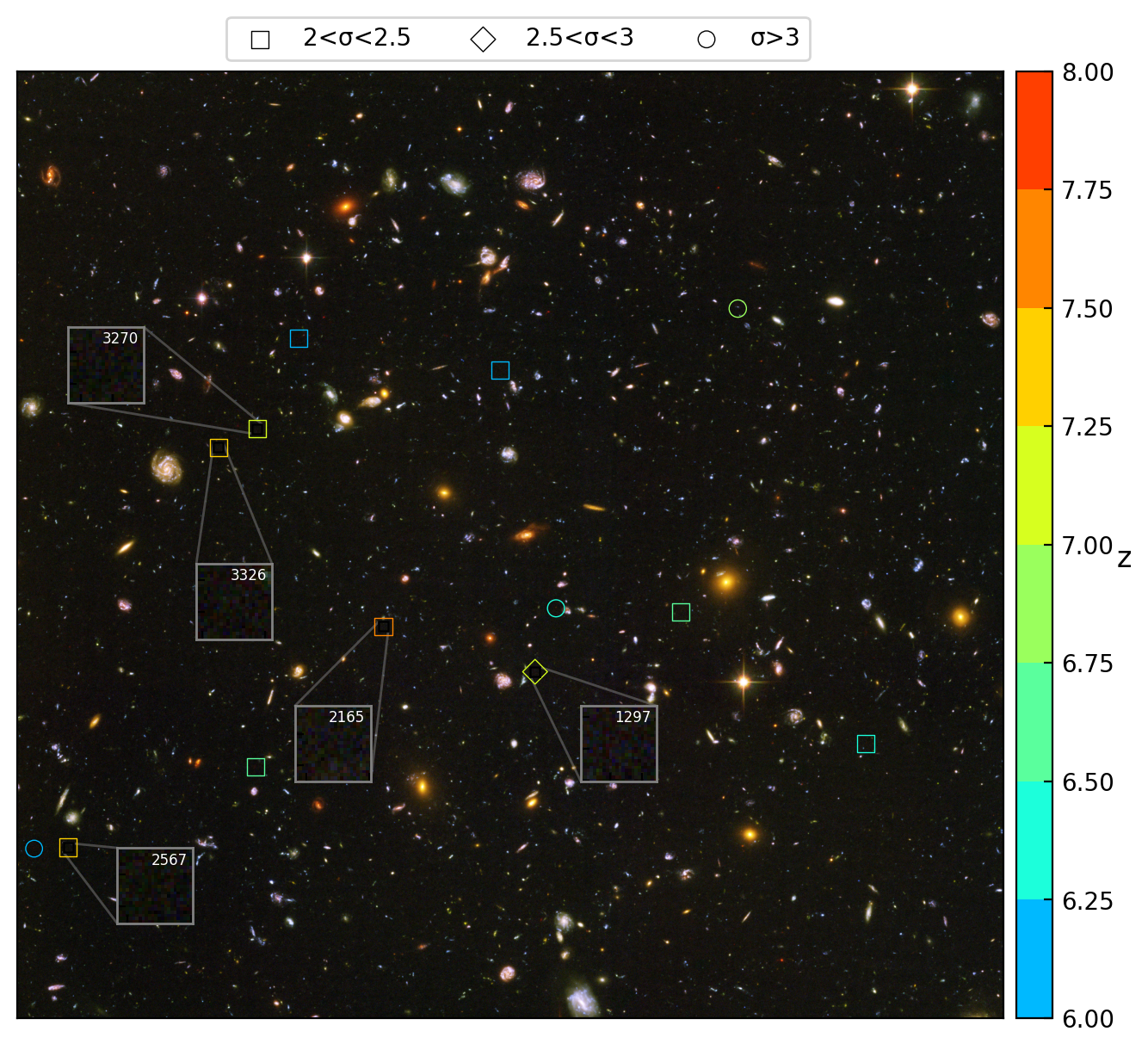}
    \caption{Angular distribution of variable candidates in the HUDF at $z\geq6$. The estimated redshift is shown by the color of the symbol with reference to the color-bar. Different symbols refer to different significance levels (see legend). We also highlight sources at $z>7$ with zoom-in thumbnails.}
    \label{fig:sky_map}
\end{figure*}

\begin{figure*}\label{fig:cutouts}
    \centering
    \includegraphics[width=0.5\textwidth]{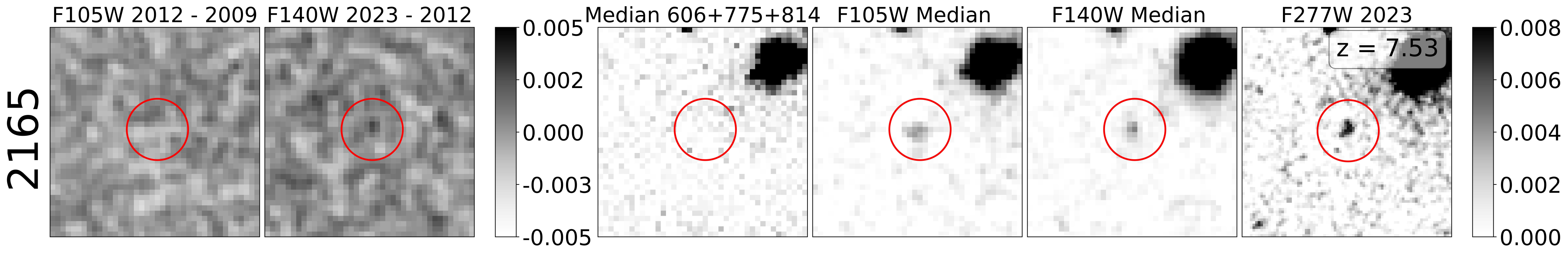} \\
    \includegraphics[width=0.5\textwidth]{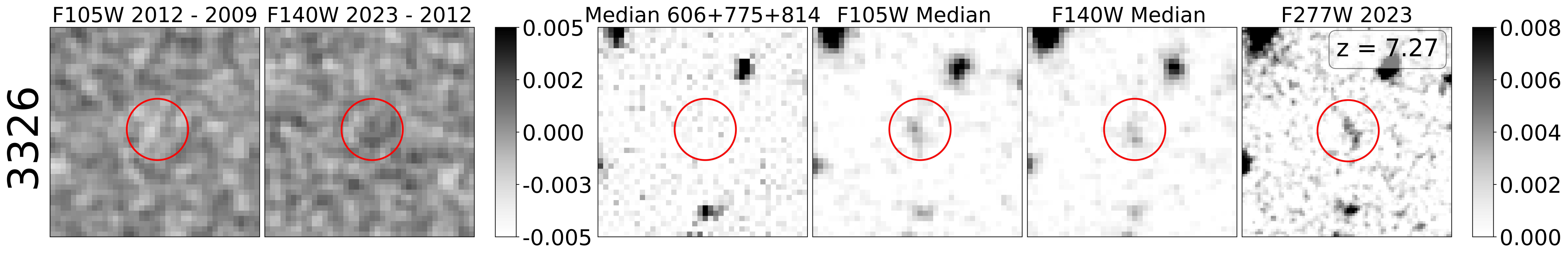} \\
    \includegraphics[width=0.5\textwidth]{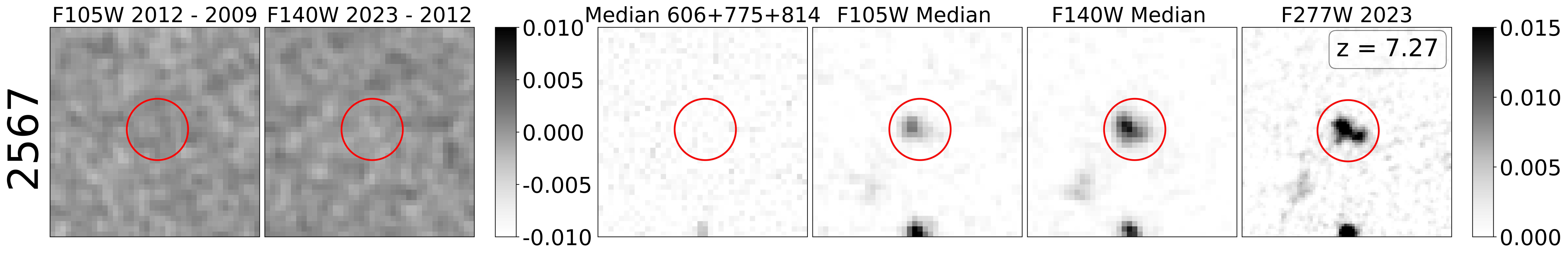} \\
    \includegraphics[width=0.5\textwidth]{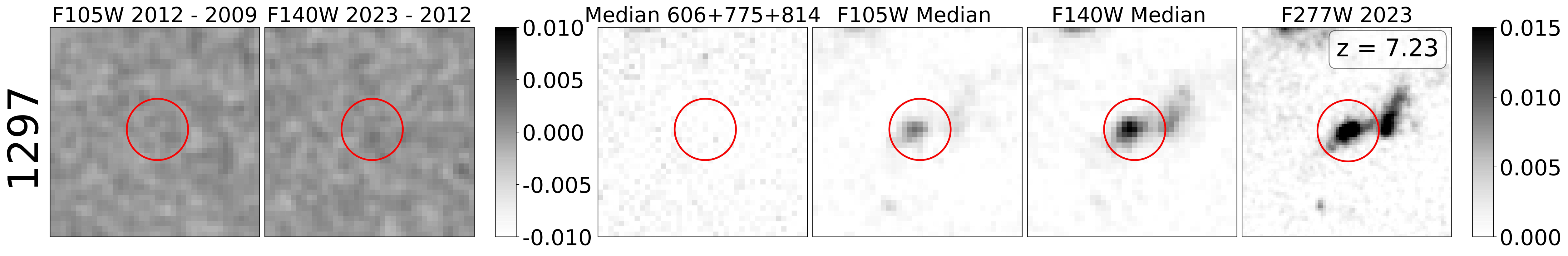} \\
    \includegraphics[width=0.5\textwidth]{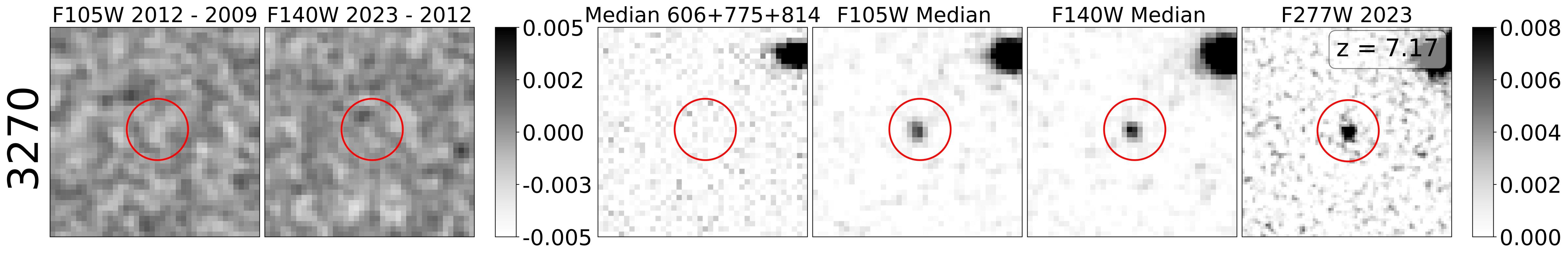} \\
    \includegraphics[width=0.5\textwidth]{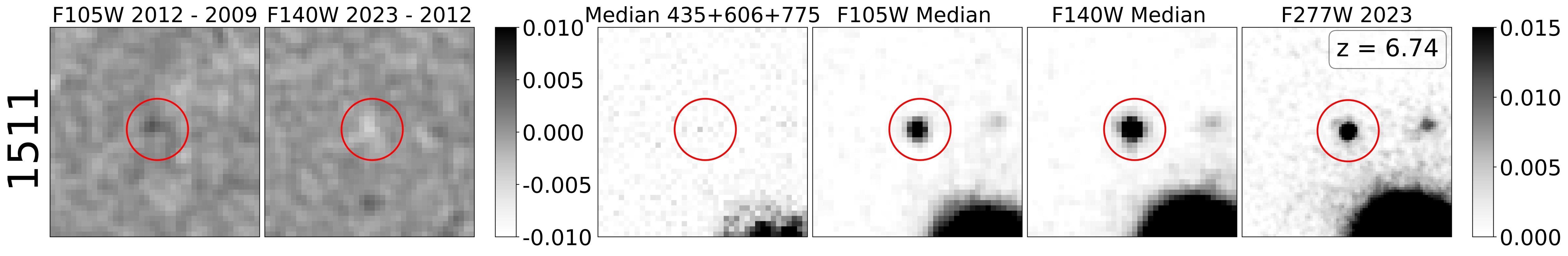} \\
    \includegraphics[width=0.5\textwidth]{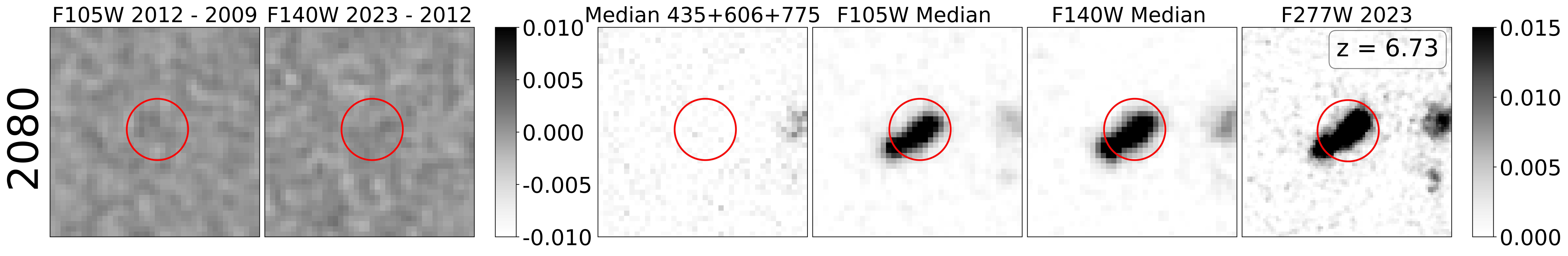} \\
    \includegraphics[width=0.5\textwidth]{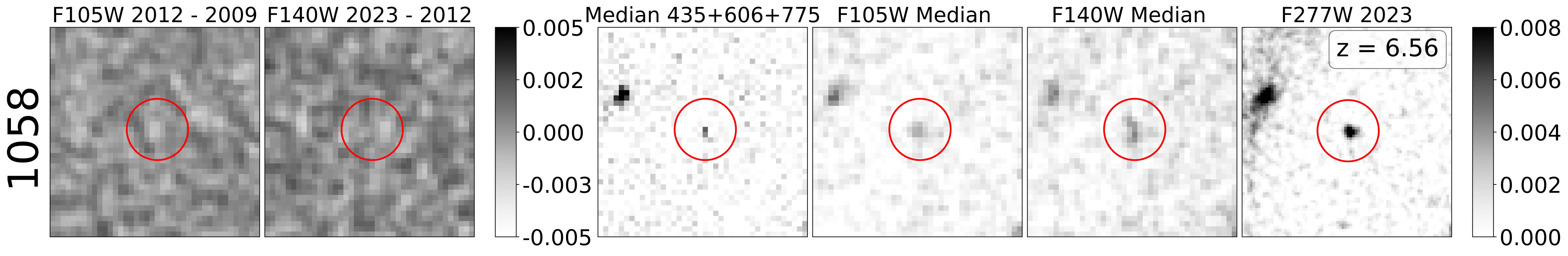} \\
    \includegraphics[width=0.5\textwidth]{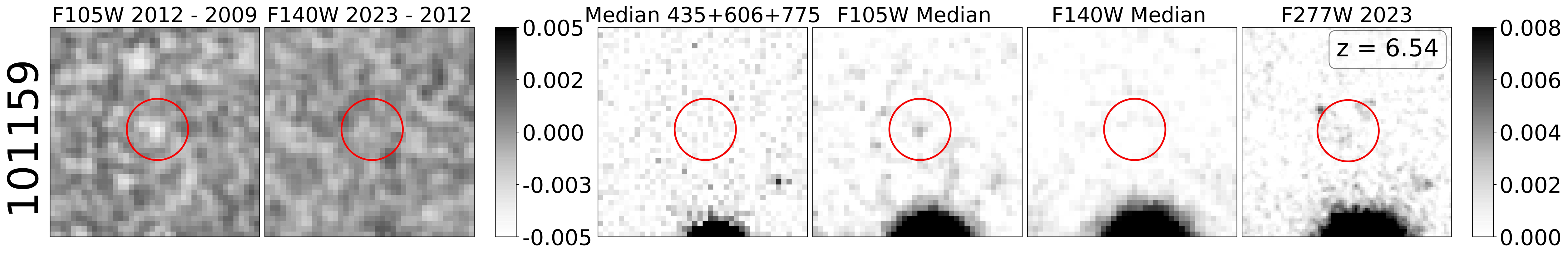} \\
    \includegraphics[width=0.5\textwidth]{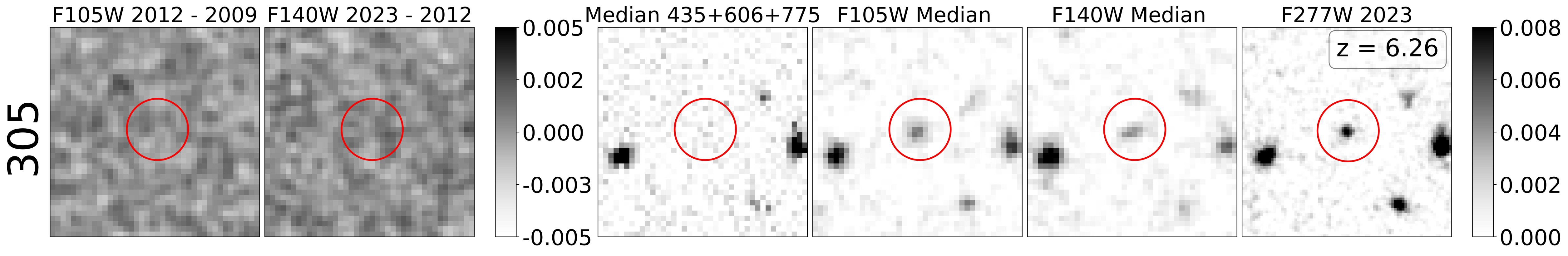} \\
    \includegraphics[width=0.5\textwidth]{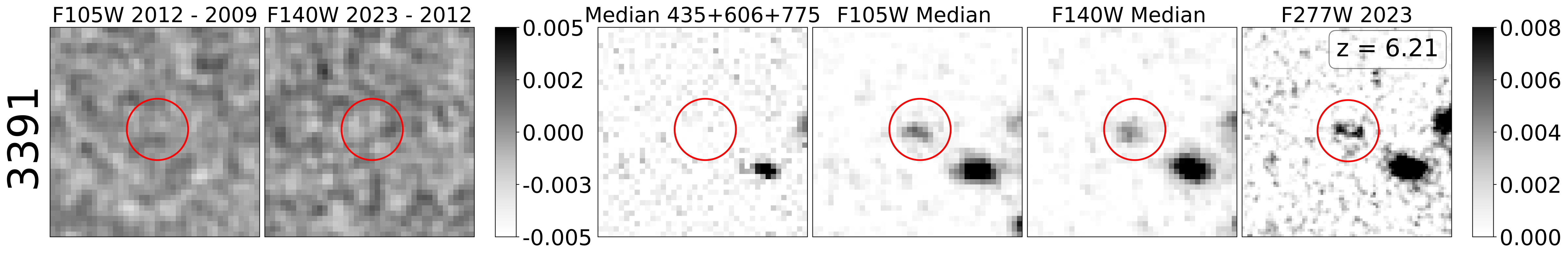} \\
    \includegraphics[width=0.5\textwidth]{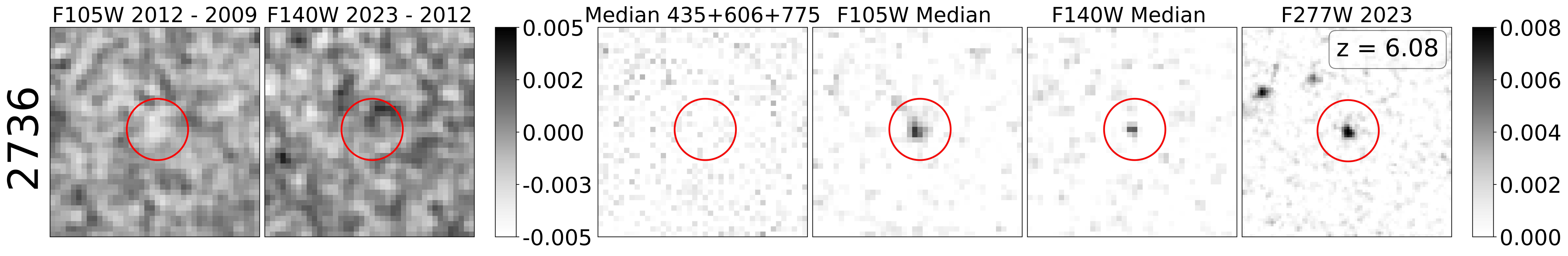} \\
    \includegraphics[width=0.5\textwidth]{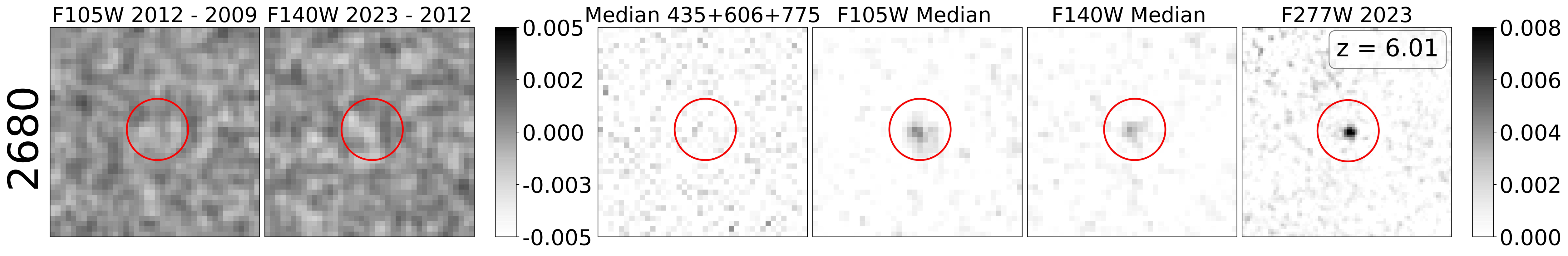} \\
    \caption{Cutout images of the sources listed in Table~\ref{tab:high_z}. In particular for each source, from left to right, we have F105W difference image, F140W difference image, median ACS image (F606W + F775W + F814W for $z > 6.9$ or F435W + F606W + F775W for $z < 6.9$) for identifying drop-outs/interlopers, the median F105W image (median stack of both epochs), median F140W image and JADES F277W image (deepest JADES filter image).}
\end{figure*}

For a visual representation of the high redshift variable sources in the field, Figures~\ref{fig:sky_map} and \ref{fig:cutouts} show their positions superimposed on the HUDF image and cutout frames in different bands used for the visual inspection, respectively. Different symbols indicate the different ranges of significance level in units of $\sigma$ as detailed in the legend, color coded according to the estimated redshift. Zoom-in images of the highest redshift ($z>7$) variable candidates are presented in small inset panels in Figure~\ref{fig:sky_map}.

\begin{table*}
    \hspace*{-2.35cm}
    \resizebox{1.12\textwidth}{!}{
        \begin{tabular}{c|cc|cc|cc|cc|cccc|c}
        \toprule
        ID & RA & Dec & $m_{1,F105W}$ & $m_{2,F105W}$ & $m_{1,F160W}$ & $m_{2,F160W}$ & $m_{1,F140W}$ & $m_{2,F140W}$  & $\sigma_{F105W}$\textsuperscript{1} & $\sigma_{F160W}$\textsuperscript{1} & $\sigma_{F140W}$\textsuperscript{2} & $\sigma_{\rm TOT}$ & Redshift \\ 
        & (deg) & (deg) & (mag) & (mag) & (mag) & (mag) & (mag) & (mag) & & & & \\
        \midrule
2165   & 53.16823 & -27.78223 &  $30.00 \pm 0.19$  &  $30.44 \pm 0.17$  &  $30.53 \pm 0.24$  &  $30.51 \pm 0.35$  &  $30.74 \pm 0.28$  & $30.05 \pm 0.16$ & -1.58  & -0.02 & \textbf{2.03}                     & 2.57  & $7.53^{+0.46}_{-0.29}$ \\ 
3326   & 53.16744 & -27.77195 &  $30.08 \pm 0.21$  &  $30.81 \pm 0.24$  &  $30.86 \pm 0.32$  &  $31.18 \pm 0.64$  &  $30.88 \pm 0.32$  & $30.16 \pm 0.17$ & \textbf{-2.31}  & -0.49 & 1.85                     & 3.0   & $7.274$\textsuperscript{b} \\ 
2567   & 53.18626 & -27.77897 &  $29.31 \pm 0.13$  &  $29.26 \pm 0.06$  &  $28.79 \pm 0.05$  &  $28.81 \pm 0.07$  &  $28.70 \pm 0.05$  & $28.85 \pm 0.06$ & 0.49   & -0.13 & \textbf{-2.20}                    & 2.26  & $7.27^{+0.11}_{-0.75}$ \\ 
1297   & 53.16481 & -27.7882  &  $29.18 \pm 0.09$  &  $29.19 \pm 0.05$  &  $28.71 \pm 0.05$  &  $28.52 \pm 0.06$  &  $28.70 \pm 0.04$  & $28.60 \pm 0.04$ & 0.16   & \textbf{2.76}  & 1.52                     & 3.16  & $7.227$\textsuperscript{c} \\ 
3270   & 53.16553 & -27.77259 &  $29.57 \pm 0.13$  &  $30.02 \pm 0.12$  &  $29.60 \pm 0.10$  &  $30.14 \pm 0.24$  &  $29.67 \pm 0.11$  & $29.76 \pm 0.12$ & \textbf{-2.34}  & \textbf{-2.06} & -0.6            & 3.18  & $7.17^{+0.09}_{-0.07}$ \\  \hline 
1511\textsuperscript{a}   & 53.16192 & -27.78699 &  $28.48 \pm 0.06$  & $28.22 \pm  0.02$  &  $28.05 \pm 0.02$  &  $27.77 \pm 0.03$  &  $27.83 \pm 0.02$  & $27.99 \pm 0.02$ & \textbf{4.58}   & \textbf{7.41}  & \textbf{-4.8}   & 9.95  & $6.74^{+0.04}_{-0.07}$ \\ 
2080   & 53.17732 & -27.78238 &  $28.21 \pm 0.04$  &  $28.14 \pm 0.02$  &  $28.04 \pm 0.03$  &  $28.07 \pm 0.04$  &  $28.06 \pm 0.03$  & $28.01 \pm 0.03$ & \textbf{2.15}    & -0.51 & 1.36                    & 2.59  & $6.73^{+0.05}_{-0.03}$ \\ 
1058   & 53.15794 & -27.79093 &  $30.03 \pm 0.19$  &  $30.73 \pm 0.22$  &  $29.96 \pm 0.14$  &  $30.28 \pm 0.27$  &  $30..6 \pm 0.15$  & $30.14 \pm 0.18$ & \textbf{-2.40}  & -1.07 & -0.36                    & 2.65  & $6.56^{+0.001}_{-6.37}$ \\ 
101159\textsuperscript{b} & 53.16052 & -27.78593 &  $28.96 \pm 0.37$  &  $\gtrsim 30.1$    &  $28.96 \pm 0.37$  &  $\gtrsim 30.1$    &  $\gtrsim 29.9$    & $\gtrsim 29.9$   & ... & ... & ....                         & ...  & $6.54^{+2.45}_{-2.59}$ \\ 
305    & 53.1564  & -27.80044 &  $30.02 \pm 0.19$  &  $29.96 \pm 0.11$  &  $30.51 \pm 0.24$  &  $29.87 \pm 0.19$  &  $30.23 \pm 0.18$  & $30.17 \pm 0.18$ & 0.48   & \textbf{2.08}  & 0.27                     & 2.15  & $6.26^{+0.19}_{-0.02}$ \\   
3391   & 53.16101 & -27.77124 &  $29.92 \pm 0.18$  &  $29.76 \pm 0.09$  &  $29.78 \pm 0.12$  &  $30.50 \pm 0.33$  &  $29.89 \pm 0.13$  & $30.00 \pm 0.15$ & 0.95   & \textbf{-2.01} & -0.52                    & 2.29  & $6.21^{+0.31}_{-0.05}$ \\  
2736   & 53.18742 & -27.77797 &  $29.44 \pm 0.15$  &  $30.53 \pm 0.25$  &  $31.04 \pm 0.50$  &  $31.72 \pm 1.16$  &  $30.93 \pm 0.46$  & $30.37 \pm 0.29$ & \textbf{-3.67}  & 0.0   & 0.97                     & 3.79  & $6.08^{+0.03}_{-0.03}$ \\  
2680   & 53.15549 & -27.77824 &  $29.97 \pm 0.18$  &  $30.33 \pm 0.15$  &  $31.43 \pm 0.53$  &  $30.33 \pm 0.28$  &  $30.09 \pm 0.15$  & $30.99 \pm 0.37$ & -1.36  & 1.78  & \textbf{-2.31}                    & 3.22  & $6.01^{+0.05}_{-0.06}$ \\  
        \bottomrule
        \end{tabular}
        }
    \caption{General properties of the $z>6$ variable sources at $>2\sigma$ significance (bold text) combining the three different filters. \textsuperscript{a}This source corresponds to source 1052123 in Paper I. 
    \textsuperscript{b}This source is taken from \citet{Hayes24} and was only detected via image subtraction method as described in the text. 
    \textsuperscript{1}Comparing epochs taken in 2008-9 vs 2012. \textsuperscript{2}Comparing epochs taken in 2012 vs 2023.}
    \label{tab:high_z}
\end{table*}

\subsection{Co-moving number density of SMBHs}\label{sec:n_smbh}

Next, assuming the detected variable sources are AGN powered by accreting SMBHs, we use the numbers of variable sources to estimate the co-moving number density of SMBHs, $n_{\rm SMBH}(z)$. As discussed in Paper I, we expect variable sources detected at high redshifts to be more likely to be AGN rather than supernovae or other stellar transients. Also, our focus on galactic nuclei will also tend to reduce supernova interlopers in our AGN sample. Very high redshift sources, i.e. $z\gtrsim6$, where the lower 1 sigma uncertainty in redshift is compatible with a low redshift interloper scenario, i.e., $z_{phot}-z_{low}\gtrsim2$, are discarded from the number density estimate.

\begin{figure*}
	\includegraphics[width=0.9\textwidth]{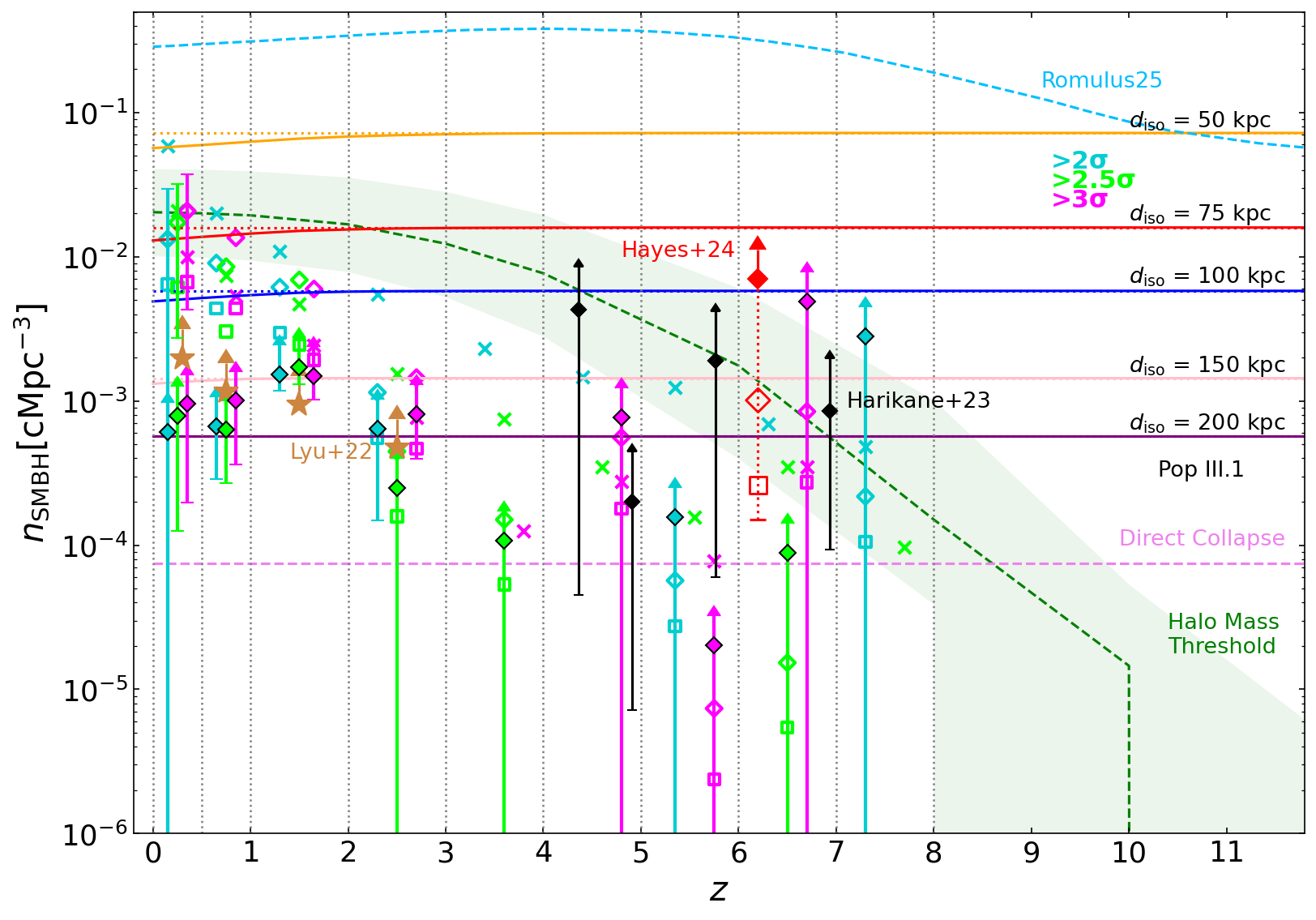}
    \caption{Evolution of the co-moving number density of SMBHs, $n_{\rm SMBH}$, versus redshift, $z$, in intervals marked by the vertical dotted lines. We report constraints based on candidate AGN detected as $2, 2.5, 3\sigma$ variability significance thresholds with turquoise, green, magenta symbols, respectively. In each bin, we report number densities based on raw counts (crosses), raw counts corrected for expected false positives (open squares), variability incompleteness corrected counts (open diamonds), and luminosity incompleteness corrected counts (solid diamonds) (see text).
    Pop III.1 SMBH seeding models \citep{Banik.2019,Singh.2023} with isolation distance parameters, $d_{\rm iso}$, from $50$ to $200$ kpc (proper distance) are shown by the colored solid lines. At low redshifts these decrease compared to the maximum value attained (dotted lines) due to mergers. The green dashed line shows the SMBH seeding assumed in \protect\cite{Vogelsberger.2014} based on a Halo Mass Threshold (HMT) above $7.1\times10^{10}\:M_\odot$ (shaded region shows a factor of two variation in this mass scale). The SMBH abundance achieved in a simulation of SMBH seeding via Direct Collapse \citep{Chon.2016} is shown by the pink dashed line. The dashed light blue line shows the BH number density as obtained in the Romulus25 simulation based on the physical condition of baryon particles \citep{Tremmel17}.
    }
    \label{fig:n_smbh}
\end{figure*}

Figure~\ref{fig:n_smbh} and Table~\ref{tab:z} present the information related to this analysis, starting with the adopted redshift intervals and the raw number of SMBHs in each interval, $N_{\rm SMBH,raw}$ (for each of the 2, 2.5 and $3\sigma$ significance levels). Next we list the total number of galaxies analysed, $N_{\rm gal}$, which is used to estimate the number of false positives, $N_{\rm fp}$, that would result at each significance level. Note that there are three chances for selection, so that the fractions of false positives $f_{\rm fp}$, assuming uncorrelated Gaussian distributions in the 3 filters, is given by
\begin{equation}\label{eq:false_pos}
    f_{\rm fp} = 1 - (1-f_{\rm fp,\sigma})^3,
\end{equation}
where $f_{\rm fp,\sigma}$ is the fraction of false positive detections expected according to a Gaussian distribution at a given $\sigma$ level. Evaluating this equation at the 2, 2.5 and $3\sigma$ levels, we find $f_{\rm fp}$ values of 0.13, 0.037 and 0.008, respectively. The final estimate of the number of SMBHs in each interval is then $N_{\rm SMBH,raw-fp}=N_{\rm SMBH,raw}-N_{\rm fp}$. 
Then, using the co-moving volume of the HUDF footprint projected across each redshift interval, $V$, we evaluate the co-moving number densities $n_{\rm SMBH,raw}$ and $n_{\rm SMBH,raw-fp}$.

We consider $n_{\rm SMBH,raw-fp}$ to be the most reliable, direct estimate of a lower limit to the co-moving number density of SMBHs. These values are plotted as open squares in Figure~\ref{fig:n_smbh}. We see that in the lowest redshift bin from $z=0$ to 0.5 that $n_{\rm SMBH,raw-fp}\simeq 6\times 10^{-3}\:{\rm cMpc}^{-3}$ regardless of the $\sigma$ threshold. We note that these values are a few times higher than the previous $z\sim0$ estimate of $n_{\rm SMBH}$ estimated by \citet{Vika.2009} and \citet{Banik.2019}. At the highest redshifts, i.e., $z\sim 7 - 8$, we find $n_{\rm SMBH,raw-fp}\sim  10^{-4}\:{\rm cMpc}^{-3}$. In between these redshift extremes we derive smoothly declining intermediate values of $n_{\rm SMBH,raw-fp}$ as redshift increases.

However, there are at least two factors causing the estimate of $n_{\rm SMBH,raw-fp}$ to be an incomplete census of SMBHs. First, we do not expect all AGN to vary during a given time interval, especially if that interval becomes relatively short in the AGN rest frame. To estimate a variability incompleteness factor, $F_{\rm var}$, we compare our sources with the sample of 31 known AGNs in the GOODS-S field, as compiled by \cite{Lyu.2022}. These AGN were identified through various methods, including mid-IR colours (4 AGN), X-ray luminosities (7 AGN), radio loudness (7 AGN), optical spectroscopy (1 AGN), and, in some cases, variability. Note, several AGNs were identified by more than one diagnostic technique. Table~\ref{tab:lyu} reports the recovery fraction, $f_{\rm AGN}$, of these 31 known AGN in our 2, 2.5 and 3$\sigma$ significance samples, i.e., 15/31, 11/31 and 10/31, respectively (note, the $3\sigma$ result includes the difference image selected sources). As a simple empirical method, we adopt the inverse of these recovery fractions as our variability incompleteness factors, $F_{\rm var}$, i.e., taking values 2.07, 2.81 and 3.10. We note that these factors have significant Poisson uncertainties, which we include in the analysis. We also note that the \citet{Lyu.2022} sample are biased to lower redshifts, i.e., $z\lesssim 3$. We expect that variability incompleteness would tend to become larger at higher redshifts due to shorter rest-frame time baselines. On the other hand, variability amplitudes are expected to increase gradually for emission in bluer wavelengths that are those probed at higher redshifts. 

Our approach does not account for sources of incompleteness arising from the diverse AGN selection methods employed to compile the \cite{Lyu.2022} sample. For instance, dust-obscured AGNs might remain undetected in X-ray surveys, or inadequate spectral resolution could hinder precise AGN diagnostics. Additionally, host galaxy contamination, which varies with redshift and host galaxy luminosity, could influence the number of variable sources recovered via variability studies, as well as the observed inverse relationship between variability significance and AGN luminosity \citep[e.g., ][]{Suberlak21}. Nevertheless, we argue that incorporating these complexities would statistically increase the expected number of recovered AGNs in the field, of which only a fraction would be varying in the optical, consequently leading to lower recovered fractions (i.e., larger incompleteness corrections). In this scenario, our $F_{\rm var}$ would thus represent, despite considerable uncertainties, a conservative estimate. More extensive, longer term monitoring of AGN variability across a range of luminosities, wavelengths and redshifts is needed for improved estimates of $F_{\rm var}$.

After the above correction for variability incompleteness, in the low-$z$ interval we obtain co-moving number densities of SMBHs of $n_{\rm SMBH,var}\simeq  2 \times 10^{-2}\:{\rm cMpc}^{-3}$ (open diamonds in Fig.~\ref{fig:n_smbh}), with little sensitivity to the choice of significance level. This may indicate our estimate of the variability incompleteness factor as a function of significance level is reasonable. The smallest uncertainties result from the $2.5\sigma$ selected sample, formally yielding $n_{\rm SMBH,var} = 1.7\pm 1.4 \times 10^{-2}\:{\rm cMpc}^{-3}$. We regard this estimate as our best measure of the true local number density of SMBHs. However, it is likely to still be a lower limit if there are very faint AGN whose flux and flux variations would not be detected by the HST observations. 
At the highest redshifts, $z= 7-8$, we find $n_{\rm SMBH,var}\simeq  2 \times 10^{-4}\:{\rm cMpc}^{-3}$. Given that we have adopted a redshift independent variability correction factor, the redshift dependence of $n_{\rm SMBH,var}$ is the same as that of $n_{\rm SMBH,raw-fp}$.

Finally, we correct for ``luminosity incompleteness'' via a luminosity correction factor, $F_{\rm lum}$, i.e., $n_{\rm SMBH,lum} = F_{\rm lum} n_{\rm SMBH,var}$. In particular, at high redshift we are only able to detect relatively bright AGN (e.g., at $z\sim6-7$ this corresponds at $M_{\rm UV}\simeq-18.6$), so this correction factor will boost the inferred number density of SMBHs. Following the method of Paper I and \citet{Harikane.2023agn}, who detected AGN via broad emission lines, we extrapolate from our detected AGN down an assumed luminosity function to a level of $M_{\rm UV}=-17\:$mag. We note that this is an arbitrary level, but by adopting this choice we are able to make a fair comparison to these previous studies. The value of the luminosity correction factor, $F_{\rm lum}$, generally becomes larger with redshift, but also depends on the form of the UV luminosity function (UVLF) we assume. Specifically, we integrate the double power-law fits detailed in \cite{Finkelstein22} from redshift $z \sim 3$ to 9. At lower redshift, we integrate Schechter function fits to the UVLF from UVCANDELS and GALEX survey programs \citep{Arnouts2005,Sun24}. At $z\lesssim4$ we see that $F_{\rm lum}$ becomes smaller than unity as our detection limit becomes fainter than the UVLF lower bound of $M_{\rm UV} = -17$. Note that in the low~$z$ interval, i.e., at $z=0-0.5$, our sensitivity corresponds to an absolute magnitude of about $M_{\rm UV}=-8.5$ assuming an average redshift of $z\sim0.25$.  This explains why we consider $n_{\rm SMBH,var}$ as our best estimate at low redshifts. In Table~\ref{tab:z} we report the values of $F_{\rm lum}$ used in each redshift interval.

After the above correction factor for luminosity incompleteness, in the high-redshift regime ($z=7-8$) we find $n_{\rm SMBH,lum}\sim 3\pm 6 \times 10^{-3}\:{\rm cMpc}^{-3}$. We caution that this result is consistent with a null density and involves a relatively large luminosity function correction factor of $F_{\rm lum}\simeq 13$ and that the systematic uncertainty associated with this correction factor is not included in the above estimate. The redshift 6 to 7 interval has similarly high values of $n_{\rm SMBH,lum}$ (although the sample based on $2\sigma$ significance does not yield a number of sources greater than those expected from false positives). At intermediate redshifts, $z=3 - 6$ we notice that $n_{\rm SMBH,lum}$ takes smaller values $\sim 10^{-3} - 10^{-4}\:{\rm cMpc}^{-3}$. We do not expect such a rapid decline in the true number density of SMBHs, so this result may either indicate that a larger fraction of SMBHs were brighter than $M_{\rm UV}=-17\:$mag at $z>6$ or that the values at high redshift are somewhat overestimated via their luminosity function correction factor. At low redshifts, we see $n_{\rm SMBH,lum}$ rises again to values of a $\lesssim 10^{-3}\:{\rm cMpc}^{-3}$. However, at $z=0-0.5$ this is a factor of about 20 smaller than the number we infer after variability completeness correction. This is simply a reflection of the fact that the observations are able to detect AGN that are much fainter than the arbitrary limit of $M_{\rm UV}=-17\:$mag.   

\begin{figure*}
	\includegraphics[width=0.9\textwidth]{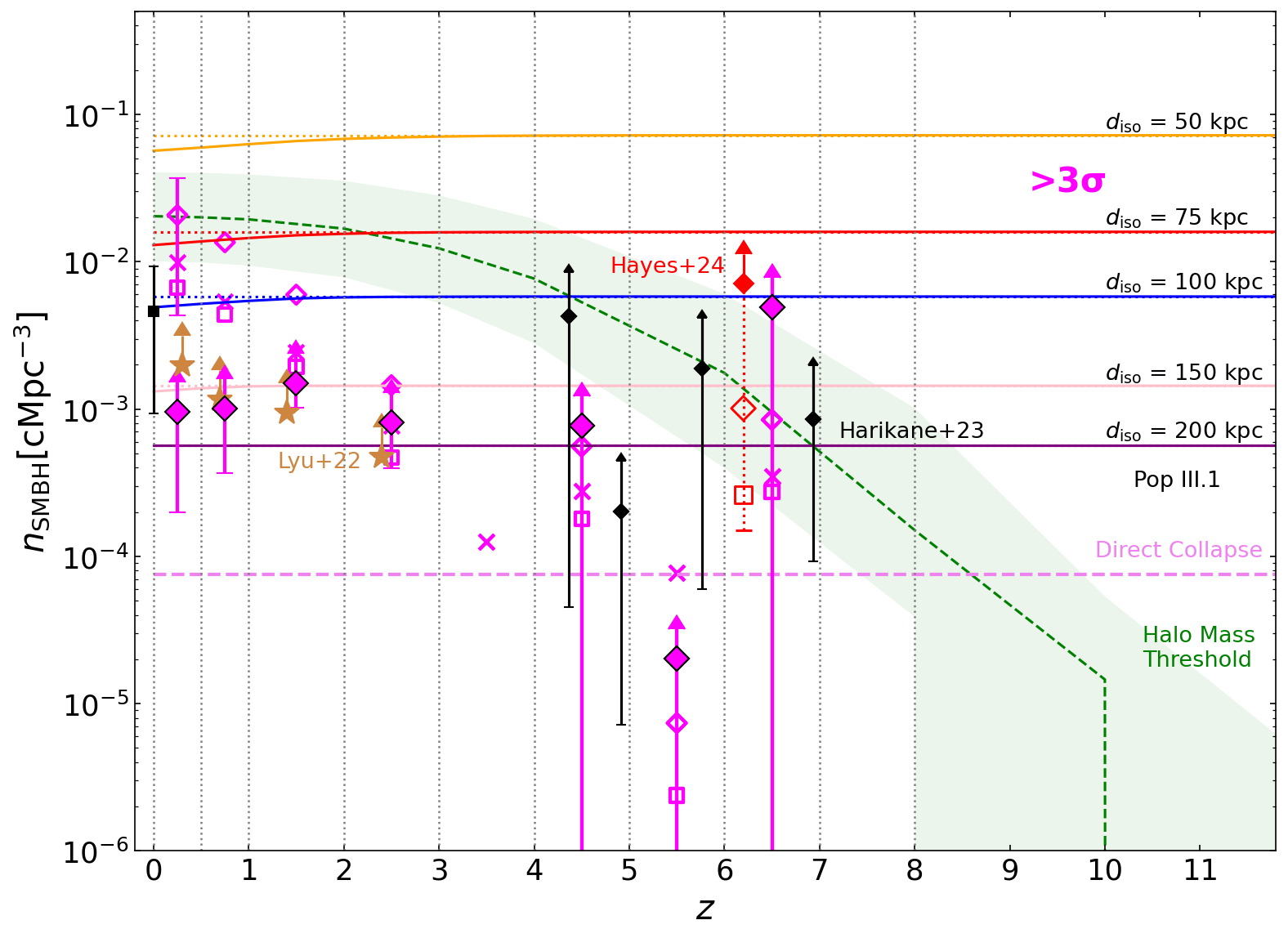}
    \caption{As Fig.~\ref{fig:n_smbh}, but now only showing $n_{\rm SMBH}$ estimated from $3\sigma$ variables. Various other observational constraints are displayed: the red diamond shows the observational constraint derived in Paper I \citep{Hayes24} based on three detected AGN between $z=6$ and 7. Broad emission line sources at $z=4-7$ \citep{Harikane.2023agn} are shown with the black diamonds, which use similar luminosity correction factors as our study. Brown stars indicate the number densities from the sample of 31 previously known AGN in the HUDF \citep{Lyu.2022}. The black square shows a previous estimate of the local ($z=0$) co-moving number density of SMBHs \citep{Banik.2019}.
    }
    \label{fig:n_smbh_3sig}
\end{figure*}

In Figure~\ref{fig:n_smbh_3sig} we show our $3\sigma$ results for $n_{\rm SMBH,var}$ and $n_{\rm SMBH,lum}$, along with several other estimates of $n_{\rm SMBH}$. As mentioned, the previous estimate of $n_{\rm SMBH}(z=0)$ of $\sim 5\times 10^{-3}\:{\rm cMpc}^{-3}$ \citep{Banik.2019} and $\sim 9\times 10^{-3}\:{\rm cMpc}^{-3}$ \citep{Vika.2009} now appear to be superseded by about a factor of a few with our HUDF variability study estimate of $n_{\rm SMBH,var}$. Similarly, we see that our values of $n_{\rm SMBH,var}$ out to $z=3$ are about 2 times higher than the values of $n_{\rm SMBH}$ implied by the 31 previously known AGN in the HUDF \citep{Lyu.2022}. 

At high redshifts we find consistency with our result from Paper I based on three detected AGN candidates between $z=6$ and 7. Our results for $n_{\rm SMBH,lum}$ at $z>6$ exceed those inferred from the JWST selected sample of \citet{Harikane.2023agn} by about a factor of 5 
From $z=4$ to 6 our estimates of $n_{\rm SMBH,lum}$ are instead about 5 times lower than those of \citet{Harikane.2023agn}. For all these high-$z$ results, there are large uncertainties due to the small number of directly detected sources. Indeed our $2\sigma$ selected sample at $z>7$ has the smallest Poisson errors. 

\begin{table*}
    \centering
        \caption{Number of variable sources retrieved in each filter and the combined numbers. 
        We also report the fraction of the 31 known AGN from the \citet{Lyu.2022} sample, $f_{\rm AGN}$, that are recovered. The inverse of this fraction is used as the variability completeness correction factor, $F_{\rm var}=f_{\rm AGN}^{-1}$.\label{tab:lyu}}
        \hspace*{-1cm}
        \begin{tabular}{ccccccc}
            \tableline
            Significance Level & \# F105W & \# F140W & \# F160W & \# combined & $f_{\rm AGN}$  & $F_{\rm var}$ \\ \tableline
            $>2\sigma$ & 222 & 191 & 185 & 521 & 15/31 &  2.07 \\ \tableline
            $>2.5\sigma$ & 92 & 67 & 68 & 188 & 11/31 &  2.81 \\ \tableline
            $>3\sigma$ & 43 & 34 & 39 & 99+10 & 10/31 &  3.10 \\ \tableline
        \end{tabular}
\end{table*}

\begin{center}
    \begin{table*}
        \caption{Census of SMBHs across the Universe in various redshift intervals out to $z=9$. In each interval we report the raw number of variable sources, $N_{\rm SMBH,raw}$, at 2, 2.5 and 3$\sigma$ significance (see text), based on analysis of a total number of galaxies, $N_{\rm gal}$, in that interval. Next, we list the number of expected false positives, $N_{\rm fp}$. The excess number of variables above this false positive level is listed as $N_{\rm SMBH,raw-fp}$. The co-moving volume for each interval is listed next, which is then used to evaluate co-moving number densities based on raw, $n_{\rm SMBH,raw}$, and false positive corrected, $n_{\rm SMBH,raw-fp}$, counts. For the latter we also list its uncertainty due to Poisson counting statistics, $\delta n_{\rm SMBH,raw-fp}$. Next we list the variability incompleteness corrected estimate $n_{\rm SMBH,var}$ and its uncertainty (see text). Next we list the luminosity incompleteness correction factor, $F_{\rm lum}$ (see text), which is then used to estimate $n_{\rm SMBH,lum}$ and its uncertainty.\label{tab:z}}
        \hspace*{-2cm}
        \resizebox{1.1\textwidth}{!}{
            \begin{tabular}{c|ccccccccccccccc}
                \tableline
                Redshift & $\sigma$ & $N_{\rm SMBH,raw}$ & $N_{\rm gal}$ & $N_{\rm fp}$ & $N_{\rm SMBH,raw-fp}$ & $V$ & $n_{\rm SMBH,raw}$ & $n_{\rm SMBH,raw-fp}$ & $\delta n_{\rm SMBH,raw-fp}$ & $n_{\rm SMBH,var}$ & $\delta n_{\rm SMBH,var}$ & $F_{\rm lum}$ & $n_{\rm SMBH,lum}$ & $\delta n_{\rm SMBH,lum}$ \\ 
                & & & & & & ($10^3 \rm cMpc^{3}$) & ($10^{-3}\rm cMpc^{-3}$) & ($10^{-3}\rm cMpc^{-3}$) & ($10^{-3}\rm cMpc^{-3}$) & ($10^{-3}\rm cMpc^{-3}$) & ($10^{-3}\rm cMpc^{-3}$) & & ($10^{-3}\rm cMpc^{-3}$) & ($10^{-3}\rm cMpc^{-3}$) & \\ \tableline
                \multirow{3}{*}{0-0.5} & 2   &  55 &     &  52.1 & 27.3 &      & 58.5 &  6.4 &  7.7 & 13.2 & 16.2 &       &  0.6  &  0.7 \\
                                       & 2.5 &  21 & 400 &  14.7 & 16.1 &  1.0 & 20.8 &  6.1 &  4.8 & 17.2 & 14.5 & 0.046 &  0.8  &  0.7 \\ 
                                       & 3   &  10 &     &   3.2 & 10.7 &      &  9.9 &  6.7 &  4.9 & 20.7 & 16.4 &       &  1.0  &  0.8 \\ \tableline
                \multirow{3}{*}{0.5-1} & 2   &  86 &     &  66.5 & 29.9 &      & 19.9 &  4.4 &  2.2 &  9.0 &  5.1 &       &  0.7  &  0.4 \\
                                       & 2.5 &  32 & 510 &  18.7 & 24.0 &  4.3 &  7.4 &  3.0 &  1.5 &  8.5 &  4.9 & 0.074 &  0.6  &  0.4 \\ 
                                       & 3   &  23 &     &   4.1 & 14.8 &      &  5.3 &  4.4 &  2.4 & 13.5 &  8.6 &       &  1.0  &  0.6 \\ \tableline
                \multirow{3}{*}{1~2}   & 2   & 162 &     & 118.2 & 50.5 &      & 11.0 &  3.0 &  0.9 &  6.1 &  2.4 &       &  1.5  &  0.4 \\
                                       & 2.5 &  70 & 907 &  33.4 & 32.5 & 14.9 &  4.7 &  2.5 &  0.7 &  6.9 &  2.9 & 0.148 &  1.7  &  0.4 \\ 
                                       & 3   &  38 &     &   7.3 & 18.7 &      &  2.4 &  1.9 &  0.8 &  6.0 &  3.2 &       &  1.5  &  0.5 \\ \tableline
                \multirow{3}{*}{2-3}   & 2   &  92 &     &  82.4 & -0.5 &      &  5.5 &  0.6 &  0.6 &  1.1 &  1.2 &       &  0.6  &  0.5 \\
                                       & 2.5 &  26 & 632 &  23.3 &  7.7 & 16.8 &  1.5 &  0.2 &  0.3 &  0.4 &  0.9 & 0.405 &  0.3  &  0.4 \\ 
                                       & 3   &  13 &     &   5.2 & 11.9 &      &  0.8 &  0.5 &  0.3 &  1.5 &  1.0 &       &  0.8  &  0.4 \\ \tableline
                \multirow{3}{*}{3-4}   & 2   &  37 &     &  40.5 &  7.5 &      &  2.3 & -0.2 &  0.4 & -0.3 &  0.8 &       & -0.2  &  0.6 \\
                                       & 2.5 &  12 & 303 &  11.2 &  4.8 & 15.9 &  0.8 &  0.1 &  0.2 &  0.2 &  0.6 & 0.72  &  0.1  &  0.4 \\ 
                                       & 3   &   2 &     &   2.4 &  1.6 &      &  0.1 &  0.0 &  0.1 & -0.1 &  0.3 &       & -0.1  &  0.2 \\ \tableline
                \multirow{3}{*}{4-5}   & 2   &  21 &     &  22.8 &  1.2 &      &  1.5 & -0.1 &  0.3 & -0.3 &  0.7 &       & -0.4  &  0.9 \\
                                       & 2.5 &   5 & 175 &   6.4 &  1.6 & 14.4 &  0.3 & -0.1 &  0.2 & -0.3 &  0.5 & 1.37  & -0.4  &  0.6 \\ 
                                       & 3   &   4 &     &   1.4 &  2.6 &      &  0.3 &  0.2 &  0.2 &  0.6 &  0.7 &       &  0.8  &  0.9 \\ \tableline
                \multirow{3}{*}{5-6}   & 2   &  16 &     &  15.6 &  3.4 &      &  1.2 &  0.0 &  0.3 &  0.1 &  0.6 &       &  0.2  &  1.8 \\
                                       & 2.5 &   2 & 120 &   4.4 &  2.6 & 12.9 &  0.2 & -0.2 &  0.1 & -0.5 &  0.4 & 2.8   & -1.5  &  1.2 \\ 
                                       & 3   &   1 &     &   1.0 &  1.0 &      &  0.1 &  0.0 &  0.1 &  0.0 &  0.2 &       &  0.0  &  0.7 \\ \tableline
                \multirow{3}{*}{6-7}   & 2   &   8 &     &  13.9 & -1.8 &      &  0.7 & -0.5 &  0.3 & -1.1 &  0.6 &       & -6.2  &  3.7 \\
                                       & 2.5 &   4 & 107 &   4.0 &  1.9 & 11.5 &  0.3 &  0.0 &  0.2 &  0.0 &  0.5 & 5.8   &  0.1  &  2.8 \\ 
                                       & 3   &   4 &     &   0.9 &  4.1 &      &  0.3 &  0.3 &  0.3 &  0.8 &  1.1 &       &  4.9  &  6.3 \\ \tableline
                \multirow{3}{*}{7-8}   & 2   &   5 &     &   3.9 &  6.7 &      &  0.5 &  0.1 &  0.2 &  0.2 &  0.5 &       &  2.8  &  6.0 \\
                                       & 2.5 &   1 &  30 &   1.1 &  2.2 & 10.9 &  0.1 &  0.0 &  0.1 &  0.0 &  0.3 & 12.9  & -0.4  &  3.5 \\ 
                                       & 3   &   0 &     &   0.3 &  1.6 &      &  0.0 &  0.0 &  0.1 & -0.1 &  0.2 &       & -1.2  &  2.2 \\ \tableline
            \end{tabular}
        }
    \end{table*}
\end{center}

\subsection{Implications for SMBH Seeding Models}

Our estimates of $n_{\rm SMBH}$ shown in Figures~\ref{fig:n_smbh} and \ref{fig:n_smbh_3sig} place strong constraints on SMBH seeding models. At high redshifts, $z>6$, we see that our estimates of $n_{\rm SMBH,lum}\gtrsim 10^{-3}\:{\rm cMpc}^{-2}$ are about a factor of 10 higher than the fiducial Halo Mass Threshold (HMT) model used in the Illustris family of simulations \citep{Vogelsberger.2014} in which dark matter halos with masses $>7\times 10^{10}\:M_\odot$ are seeded with SMBHs. The range of the green band shown in Figs.~\ref{fig:n_smbh} and \ref{fig:n_smbh_3sig} corresponds to varying this threshold mass by factors of two. Thus an HMT model with a threshold mass of $\sim3\times 10^{10}\:M_\odot$, which would be consistent with the $z=0$ constraints, would be closer to the high-$z$ data, but still low by about a factor of few.

Figures~\ref{fig:n_smbh} and \ref{fig:n_smbh_3sig} also compares our constraints with the Romulus25 simulation \citep{Tremmel17}. We see that this simulation, which seeds SMBHs using simple threshold conditions of gas properties (see \S\ref{sec:intro}), produces a high number density of SMBHs at early times, i.e., $n_{\rm SMBH} \sim 5\times 10^{-2}\:{\rm cMpc}^{-3}$ at $z\sim 12$. This rises by almost an order of magnitude by $z\simeq 4$, before then undergoing a modest decline by $z=0$. Given its high number densities of SMBHs, this model remains consistent with our observational constraints on $n_{\rm SMBH}$.

As discussed in \S\ref{sec:intro}, Direct Collapse (DC) seeing models struggle to reproduce the overall number densities of SMBHs. The most optimistic of the DC models discussed in \S\ref{sec:intro} is that of \citet{Chon16} with $n_{\rm SMBH}\sim 10^{-4}\:{\rm Mpc}^{-3}$. As shown in Figs.~\ref{fig:n_smbh} and \ref{fig:n_smbh_3sig} this is about a factor of 50 below $n_{\rm SMBH}$ inferred from our HUDF variability study. Furthermore, the later simulations of DC by \citet{Wise.2019} and \citet{2022Natur.607...48L} find values of $n_{\rm SMBH}\lesssim 10^{-6}\:{\rm Mpc}^{-3}$, i.e., at least a factor of at least $10^3$ below our high-$z$ estimate. The constraints are even more severe when considering our low-$z$ estimates of $n_{\rm SMBH}$, which have smaller uncertainties and values of $\simeq 2\times 10^{-2}\:{\rm Mpc}^{-3}$. 

A striking feature of our low-$z$ and high-$z$ estimates of $n_{\rm SMBH}$ is the relative similarity of $n_{\rm SMBH,lum}$ at high-$z$ with $n_{\rm SMBH,var}$ at low-$z$. There are number of reasons to expect that $n_{\rm SMBH,lum}(z>6)$ is a lower limit, i.e., since it only considers AGN down to $M_{\rm UV}=-17\:$mag and that variability incompleteness factors may be larger given the shorter rest-frame time intervals that are probed by the HUDF observations. This would then indicate a relatively constant evolution in $n_{\rm SMBH}$ from $z\sim8$ to $z\sim0$. Such an evolution is a key feature and prediction of the Pop III.1 SMBH seeding model of \citet{Banik.2019}, which has been further explored by \citet{Singh23} and \citet{Cammelli24.MNRAS2}. The main parameter of the Pop III.1 model is the isolation distance, $d_{\rm iso}$, needed for minihalos to be Pop III.1 sources that seed SMBHs (as opposed to Pop III.2 sources that do not). Expressed in proper distances at the time of formation, our $n_{\rm SMBH}$ results from the HUDF favour models with $d_{\rm iso}\lesssim 50\:$kpc. This would correspond to co-moving separations of $\lesssim 1\:$Mpc at typical formation redshifts of $z\sim 20$. We note that this result is consistent with that of \citet{Cammelli24.MNRAS2} of $d_{\rm iso}\lesssim 75\:$kpc based on the galaxy stellar mass function (GSMF) and the SMBH occupation fraction as a function of the stellar mass.

As discussed by \citet{2024arXiv241201828T}, more realistic physical models for $d_{\rm iso}$ based on R-type HII region expansion around supermassive Pop III.1 stars that are expected to be the direct progenitors of the SMBHs yield estimates of the
co-moving number density of SMBHs of
\begin{equation}
n_{\rm SMBH}= \frac{3}{4\pi R_R^3} \rightarrow  0.18 t_{*,10}^{-1} S_{53}^{-1}\:{\rm cMpc}^{-3},\label{eq:nlim}
\end{equation}
where $t_{*,10} = t_*/10\:$Myr, $t_*$ is the lifetime of the supermassive star, $S_{53} \equiv S/10^{53}\:{\rm s}^{-1}$, $S$ is the H-ionizing luminosity of the star, and $R_R$ is the radius of R-type HII region expansion achieved in the intergalactic medium. This estimate can be viewed as an upper limit since it has assumed
maximal close packing of the sources and ignored other contributions
to the ionizing background from Pop III.2, Pop II and AGN sources. It is interesting that our estimate of $n_{\rm SMBH}$ at low-$z$ from the HUDF is within a factor of a few of this fiducial estimate.

Similar to the analysis carried out in Paper I, a valuable test for SMBH seeding schemes consists in measuring the actual distance between candidate AGN pairs. It is helpful to remind here that the Pop III.1 model with $d_{\rm iso}\simeq 100\:$kpc in proper distance would correspond to a comoving separation of about 3~cMpc (if the relative motion by $z\sim 7$ is negligible). Also, the HUDF footprint at this epoch is $5.5$~cMpc on a side. In our sample, at $z>6$ we are left with the 25 sources listed in Table~\ref{tab:high_z} and three of them have spectroscopic redshifts available. Since we cannot assess the true distance with even one photometric redshift in a single pair (already a $\Delta_z=0.01$ at $z\sim7$ would be $\sim2.5$ cMpc, comparable to \diso), we have searched among the pairs having both spectroscopic redshift available. In this high-$z$ regime, we find none and so no constraint can be used to rule out specific scenarios. Obtaining spectroscopic redshifts of all the high-$z$ SMBH candidates is needed to achieve better constraints.

\subsection{Comparison with other AGN diagnostics}

Despite recent advances, it remains immensely challenging to obtain an accurate census of AGN at $z\gtrsim6$ at faint magnitudes. Hot dust tracers \citep[e.g., ][]{Stern2005} are shifted to the FIR, and X-ray and radio facilities are not sufficiently sensitive to identify individual faint AGN at these distances. High ionization UV emission lines (e.g., \heII~$\lambda$1640, \cIV~$\lambda$1550, \nV~$\lambda$1240) are sometimes observed in luminous targets \citep[e.g.,][]{Mainali.2017, Laporte17}, but they can be inconclusive, as these lines also form in low-metallicity starbursts \citep{Senchyna.2020, Berg.2019, Saxena20}. {\it JWST}-NIRSpec micro-shutter array (MSA) observations have delivered deep IR spectroscopy and several studies \citep[e.g.,][]{Cameron.2023, Boyett24, Hu24} could identify AGN via line flux ratios classification \citep[BPT-like diagrams,][]{BPT81}, although the targeted approach of MSA spectroscopy is subject to photometric pre-selection biases and tends to be focused more towards the brighter end of the luminosity function. In addition, NIRSpec surveys face challenges in achieving completeness at $M_{UV}\simeq-17$ to $-20$, where galaxies are $\sim 100$ times more abundant.

A number of studies \citep[e.g.,][]{Cohen.2006, Pouliasis.2019, OBrien.2024} have shown the effectiveness of using photometric monitoring with HST to identify variable sources, finding large numbers of AGN at intermediate luminosities, that evade other selection techniques (X-ray, radio, IR colours). 
In addition, as demonstrated in this series of papers, we advocate that time variability is an efficient observable for AGN identification in the high redshift Universe.

\subsection{Supernova contamination}\label{sec:sn}

Our observations are sensitive to the presence of stellar transients, especially supernovae (SNe), which may be a source of contamination in our counting of AGN. In Paper I we identified three variables sources that are supernovae, two of which are apparently hostless with indeterminable redshifts. Supernovae are expected to follow either the stellar mass distribution (thermonuclear) or star formation (SFR) distribution (core collapse).  Statistically in a galaxy these are centrally concentrated, so our approach of focussing on deep-stack centroiding in galactic nuclei, while helping to avoid off-nuclear SNe, would still be subject to some level of stellar transient contaminants. We note that we do not consider Tidal Disruption Events (TDEs), i.e, stellar disruption by a supermassive black hole, to be contaminants in our census. 

As discussed in Paper I, our {\it HST} images are not deep enough to detect ordinary core collapse SNe at $z\gtrsim3$.  We are potentially sensitive to thermonuclear SNe at these redshifts if they exist, but the delay times for the main Type Ia channels means that they should be very rare at high-$z$.  Superluminous SNe (SLSNe) are an interesting possibility, and could be detected out to $z\sim6$. However, their rates should follow the cosmic SFR density, and we expect very low numbers in the small volume probed by the HUDF. A follow-up paper in this series will focus on the full census of detected SNe candidates.

\section{Conclusions}
\label{sec:conclusions}

Through photometric monitoring conducted in three epochs (2008/2009, 2012, and 2023) we have discovered many variable sources in the Hubble Ultra Deep Field, which we consider to be likely AGN candidates tracing the presence of SMBHs. We advocate that variability searches are a highly effective and comprehensive tool for identifying AGN in deep imaging surveys. While only a fraction of AGN can be detected through their variability, the significant advantage of this approach lies in its ability to survey the entire field with no prior selection.
Unlike other methods such as radio or X-ray diagnostics, which are limited by detection thresholds, variability searches ensure that any object captured in imaging can be tested for its activity.

Historically, high-redshift luminous quasars have been known for some time \citep{Fan.2006,Mortlock.2011,Banados.2018}, though they are typically much brighter—by four magnitudes or more—than the sources we detect. More recently, there have been reports of AGN at even higher redshifts with luminosities closer to \Lstar\ \citep[e.g.,][]{Maiolino.2023agn,Harikane.2023agn,Larson.2023}. However, our survey, which was restricted to the small volume of the HUDF, has the ability to place stronger constraints than any previous work on the number densities of SMBHs at these redshifts. We have shown how the 13 variable sources we detect between $z= 6$ and 8 imply high co-moving number densities of $n_{\rm SMBH}\gtrsim 2\times 10^{-3}\:{\rm cMpc}^{-3}$. In addition, our results at low redshift make a new measurement of $n_{\rm SMBH}\sim 2\times 10^{-2}\:{\rm cMpc}^{-3}$ in this regime that is about a factor of 5 higher than previous estimates.



Our measurements constrain SMBH seeding mechanisms and their implementation in cosmological simulations. For example, our estimate of $n_{\rm SMBH}$ at $z=7-8$ is about a factor of 10 higher than expected in the fiducial halo mass threshold model used in the Illustris family of simulations \citep{Vogelsberger.2014}. Our estimates, including at $z\sim 0$, are $\sim10^2-10^4$ times larger than produced in simulations of SMBH seeding via Direct Collapse \citep{Chon.2016,Wise.2019,2022Natur.607...48L}. 

On the other hand, the relative constancy of $n_{\rm SMBH}$ from high to low $z$ confirms a prediction of the Pop III.1 SMBH seeding model of \citet{Banik.2019} and constrain its main isolation distance parameter, $d_{\rm iso}$ to be $\lesssim75\:$kpc. Such values are also consistent with recent estimates of 
SMBH occupation fractions as a function of the stellar mass \citep{Cammelli24.MNRAS2}. They are also consistent with fiducial Pop III.1 model expectations of ionizing feedback for the isolation distance  \citep{2024arXiv241201828T}.

Future papers in this series will examine the properties of the host galaxies of our detected AGN candidates (Young et al., in prep.) and carry out a systematic census for stellar transient events.

\section*{Acknowledgements} We thank Yuichi Harikane and Anna Wright for helpful discussions. We thank Peter Williams for providing the 3-color image of the HUDF used in Fig.~4 and scripts for overplotting data. V.C. thanks the BlackHoleWeather project and PI Prof. Gaspari for salary support. V.C. also acknowledges support from the Chalmers Initiative on Cosmic Origins (CICO) as a visiting Ph.D. student at Chalmers Univ. and co-funding of his former Ph.D. position at Univ. of Trieste. J.C.T. acknowledges support from ERC Advanced grant 788829 (MSTAR). A.R.Y. is supported by the Swedish National Space Agency (SNSA).  M.J.H. is supported by the Swedish Research Council (Vetenskapsr\aa{}det) and is Fellow of the Knut \& Alice Wallenberg Foundation. R.S.E. acknowledges generous financial support from the Peter and Patricia Gruber Foundation. M.J.H. and J.C.T. thank the staff in The Doors public house for service that germinated the ideas that led to this project.

\facilities{HST (WFC3)}

\section*{Data Availability}
All the {\it HST} data used in this paper can be found in MAST: \dataset[10.17909/7s5v-gz68]{http://dx.doi.org/10.17909/7s5v-gz68}.


\bibliography{a.bib}{}
\bibliographystyle{aasjournal}

\clearpage

\appendix

\section{Variability in F105W and F160W}\label{app:phot_var}

\begin{figure*}
    \centering
        \hspace{-0.5cm}
        \includegraphics[width=0.9\textwidth]{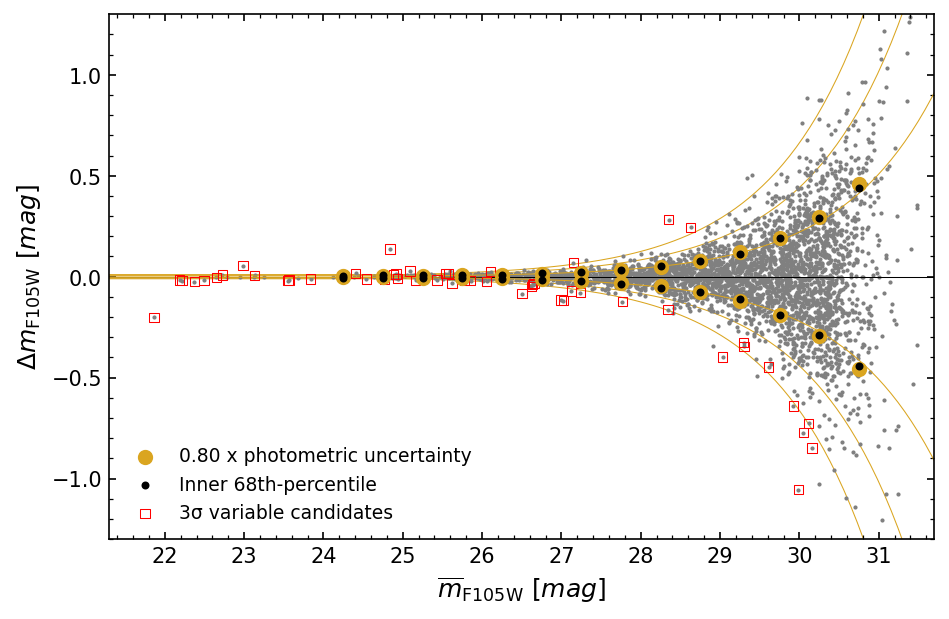}
        \includegraphics[width=0.88\textwidth]{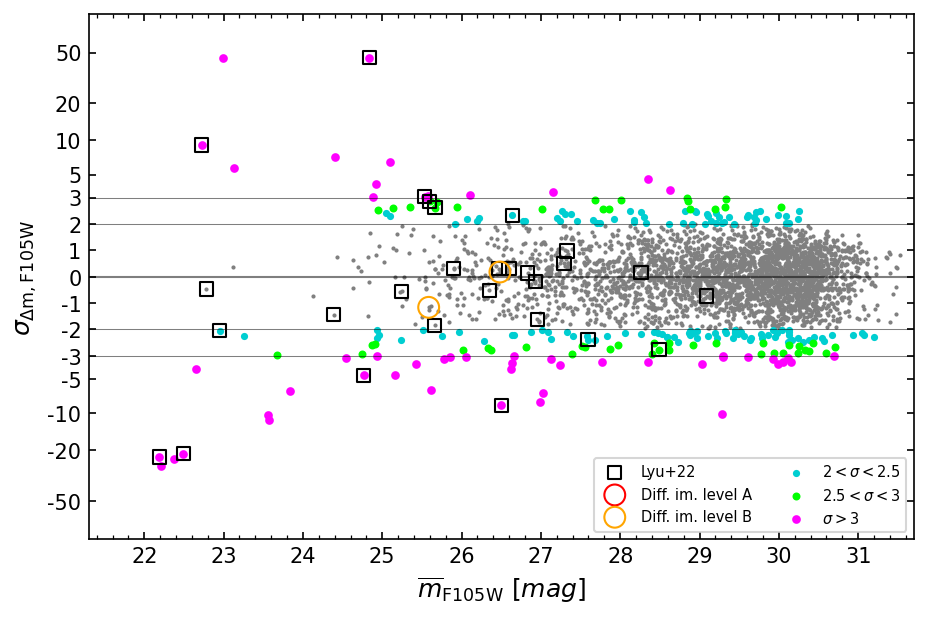}
        \caption{
        Photometric variability in the HUDF in the F105W filter from 2008-9 to 2012. Magnitude difference from epoch 1 (2008-9) to epoch 2 (2012) is plotted versus average magnitude (grey points).
        }\label{fig:delta_mag_calib_105}
\end{figure*}

\begin{figure*}
    \centering
        \hspace{-0.5cm}
        \includegraphics[width=0.9\textwidth]{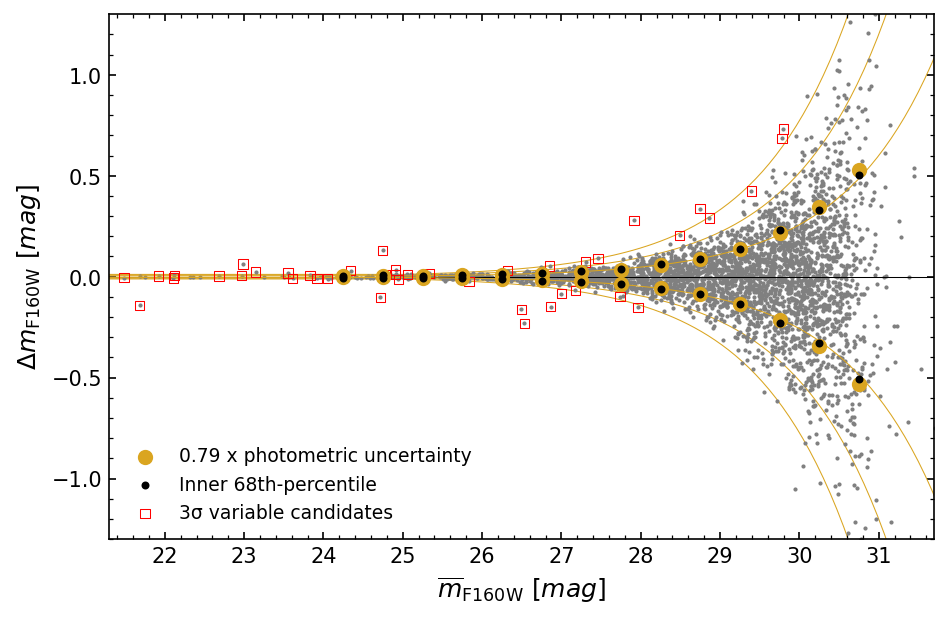}
        \includegraphics[width=0.88\textwidth]{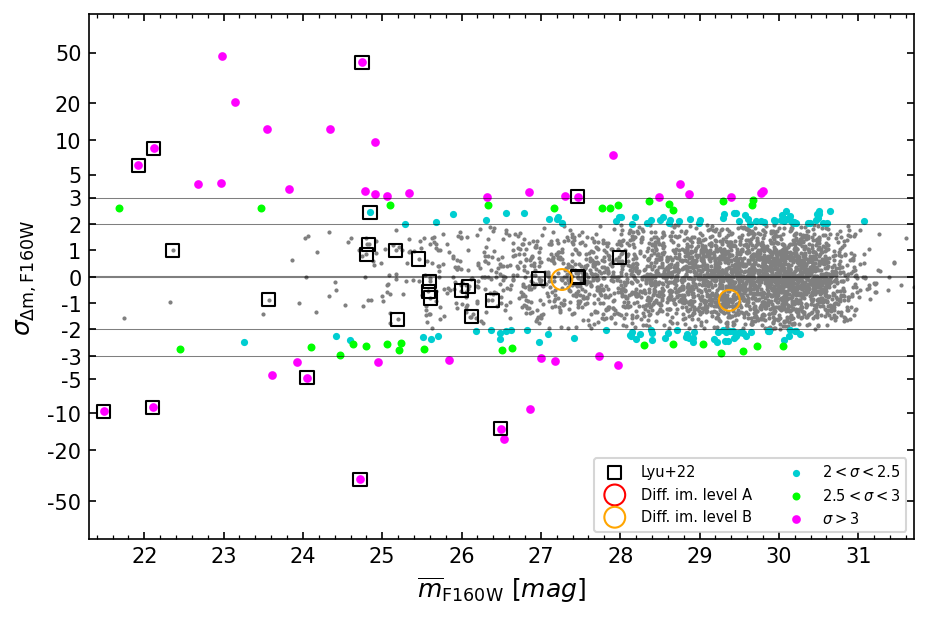}
        \caption{
        Photometric variability in the HUDF in the F160W filter from 2008-9 to 2012. Magnitude difference from epoch 1 (2008-9) to epoch 2 (2012) is plotted versus average magnitude (grey points).
        }\label{fig:delta_mag_calib_160}
\end{figure*}

Figures \ref{fig:delta_mag_calib_105} and \ref{fig:delta_mag_calib_160} report the photometric variability as a function of the mean mag in the F105W and F160W filters, as detailed in Fig.~\ref{fig:delta_mag_calib}.

\end{document}